\documentclass[11pt,a4paper]{article}
\pdfoutput=1
\usepackage{cancel}
\usepackage{enumitem}
\usepackage{etex}
\usepackage{amsthm}
\usepackage{geometry}
\usepackage{dcolumn}
\usepackage{epsf}
\usepackage{mathrsfs}
\usepackage{multirow}
\usepackage{booktabs}
\usepackage{tabularx}
\usepackage{array}
\usepackage{slashed}
\usepackage{float}
\usepackage{leftidx}
\usepackage{setspace}
\usepackage{verbatim}
\usepackage{adjustbox}
\usepackage{tikz}
\usepackage[caption=false]{subfig}
\usepackage{arydshln}
\usepackage{jheppub}
\usepackage{shuffle}
\usepackage{amsmath}
\usepackage{cases}
\usepackage{lipsum}
\usepackage{tabularx}
\usepackage{graphicx}
\usepackage{lscape}

\numberwithin{equation}{section}
\usetikzlibrary{arrows,decorations.markings,shapes.arrows,patterns,positioning}

\newcommand{\bea}{\begin{eqnarray}}
\newcommand{\eea}{\end{eqnarray}}
\newcommand{\bean}{\begin{eqnarray*}}
\newcommand{\eean}{\end{eqnarray*}}
\newcommand{\nn}{\nonumber\\}
\newcommand{\Sl}{\sum\limits}

\def\W #1{\widetilde{#1}}

\def\Label#1{\label{#1}%
  \smash{\hbox to0pt{\raise1ex\hbox{\tiny[#1]}\hss}}}
\renewcommand{\eqref}[1]{eq.~(\ref{#1})}

\newcommand{\secref}[1]{section~\ref{#1}}

\def\Sl{\sum\limits}

\newcommand{\ctobedelete}[1]{}

\allowdisplaybreaks

\title{Expansion formula of one-loop Einstein-Yang-Mills integrand}

\author[a,b]{Yi-Jian Du} \author[c]{Chongsi Xie}

\affiliation[a]{Department of Physics, School of Physics and Technology,
Wuhan University,\\
No.299 Bayi Road, Wuhan 430072, China}
\affiliation[b]{Hubei Key Laboratory of Nuclear Solid Physics, School of Physics and Technology, Wuhan University,\\
No.299 Bayi Road, Wuhan 430072, China}
\affiliation[c]{College of Science, Hainan Tropical Ocean University,\\
Sanya, Hainan 572022, P.R. China}

 \emailAdd{yijian.du@whu.edu.cn}\emailAdd{chongsi.xie@whu.edu.cn}

\date{\today}
\abstract{ 
Building upon the algebraic consistency construction of one-loop Bern-Carrasco-Johansson (BCJ) numerators for Yang-Mills (YM) and Yang-Mills-scalar (YMS) theories, we explore the expansion formula of one-loop Einstein-Yang-Mills (EYM) integrands (with a gluon loop) in terms of conventional one-loop YM integrands with quadratic propagators. We first express the EYM integrand by tree-level amplitudes according to the forward limit approach. Employing a two-step expansion strategy, the gluon-loop EYM integrand is decomposed into tree-level YM amplitudes under the forward limit, which are subsequently expanded into tree-level bi-adjoint scalar (BS) ones. We then prove that when the kinematic coefficients in both expansion steps satisfy the one-loop consistency conditions, the EYM integrand is finally expanded as a combination of YM integrands with quadratic propagators.
The coefficients in this expansion formula coincide exactly with those in the expansion formula for YMS integrands. This correspondence highlights a shared kinematic structure, providing the proper foundation for constructing BCJ numerators in both YMS and EYM theories at one loop.
}


\begin{document}
\maketitle \flushbottom

\section{Introduction}\label{sec:intro}

Colour-kinematics duality \cite{Bern:2008qj,Bern:2010ue}, proposed by Bern, Carrasco and Johansson, reveals that colour-dressed Yang-Mills (YM) amplitudes can be arranged as sums over trivalent Feynman-like graphs, each of which is associated with a colour factor and a kinematic factor. Crucially, the kinematic factors are further required to satisfy the same algebraic identities as their colour counterparts. Furthermore, it was conjectured that gravitational amplitudes in General Relativity (GR) can be obtained by replacing the colour factors in YM with their kinematic counterparts.
This duality offers profound insight into the connection between YM and GR. The central problem in understanding colour-kinematics duality is the construction of kinematic numerators (BCJ numerators) satisfying the algebraic identities.

For tree-level amplitudes, a systematic construction of BCJ numerators based on the recursive expansion formula of Yang-Mills-scalar (YMS) amplitudes was proposed in \cite{Fu:2017uzt,Chiodaroli:2017ngp,Teng:2017tbo,Du:2017kpo,Du:2017gnh}. In this approach, a colour ordered YM  amplitude is expressed in terms of doubly colour ordered YMS amplitudes which are then expanded into combinations of doubly colour ordered bi-adjoint scalar (BS) ones, according to the recursive expansion formula\footnote{Throughout this paper, all Yang–Mills amplitudes/integrands are colour ordered ones, while all YMS and BS amplitudes/integrands are doubly colour ordered ones. Further details on the relation between colour dressed and colour ordered objects can be found in, e.g., \cite{Porkert:2022efy,Du:2025yxz}.}. The coefficients obtained from this {\it two-step procedure} are polynomail functions of Lorentz contractions of external polarisation vectors and momenta. As pointed out in \cite{ref-Kiermaier,Bern:2010yg,Bjerrum-Bohr:2012kaa,Fu:2012uy,Fu:2014pya}, these coefficients effectively generate all tree level BCJ numerators satisfying algebraic identities. An important fact that follows from the double copy conjecture \cite{Bern:2008qj,Bern:2010ue} is the following. The expansion formulas which express YM and YMS amplitudes in terms of BS ones can also expresss GR and Einstein-Yang-Mills (EYM) amplitudes \cite{Fu:2017uzt,Chiodaroli:2017ngp,Teng:2017tbo,Du:2017kpo,Du:2017gnh}, in terms of colour ordered pure YM ones with the same expansion coefficients. This has been proved by using either Bern-Cachazo-Feng-Witten (BCFW) recursion \cite{Britto:2004ap,Britto:2005fq} or the Cachazo-He-Yuan (CHY) formula \cite{Cachazo:2013gna,Cachazo:2013hca,Cachazo:2013iea,Cachazo:2014nsa,Cachazo:2014xea}.

 Despite extensive studies of tree-level numerators, a general construction of one-loop BCJ numerators in YM and GR remains an open problem. The forward limit approach originating from worldsheet formalisms \cite{Mason:2013sva,He:2015yua,Cachazo:2015aol} provides a powerful tool for analysing loop integrands. Along this line, an $n$-point one-loop integrand is expressed as a sum of $(n+2)$-point tree amplitudes, where two particles carry the loop momentum $\pm\ell^{\mu}$ with opposite directions. By further expanding these tree amplitudes using the tree-level expansion formulas, one can express the one-loop YM/YMS (GR/EYM) integrands as combinations of $(n+2)$-point tree-level BS (YM) amplitudes in the forward limit \cite{He:2016mzd,He:2017spx,Geyer:2017ela,Porkert:2022efy}. However, this forward limit representation of integrands differs from the conventional one based on Feynman diagrams, in the sense that the denominators of loop propagators are linear functions of the loop momentum $\ell^{\mu}$. Consequently, the expansion coefficients obtained in this way cannot be naively used to generate one-loop BCJ numerators with respect to quadratic propagators.

Recent work \cite{Cao:2025ygu,Du:2025yxz} proposed a consistency condition stating that one‑loop BCJ numerators should remain invariant under a cyclic permutation combined with a shift of the loop momentum $\ell$. As further noted in \cite{Du:2025yxz}, the two‑step construction used at tree level can be extended to one‑loop integrands: a one‑loop Yang–Mills (YM) integrand is first expanded into Yang–Mills–scalar (YMS) integrands, and subsequently into bi‑adjoint scalar (BS) integrands. Because the coefficients in this two‑step procedure serve as a basis for one‑loop BCJ numerators, and the coefficients of the first step are independent of the loop momentum, it is natural to look for an expansion formula of YMS integrands into BS integrands whose coefficients satisfy the consistency condition. In addition to a general analysis of the existence of such coefficients, explicit expansion formulas for scalar‑loop YMS integrands with no more than three gluons and an arbitrary number of scalars were presented in \cite{Du:2025yxz}, using the localisation strategy developed in \cite{Xie:2024pro,Xie:2025utp}.

Since the two-step procedure for tree-level YM amplitudes  has been proven to be effective for GR amplitudes, one may expect a one-loop version of the two-step construction of GR integrands, which shares the same expansion coefficients as in the formula for one-loop YM integrands. Consequently, this calls for an expansion formula of EYM integrands in terms of YM integrands, with the coefficients inherited from the formula for YMS integrands. Although one can verify such EYM expansion formula by unitarity cut straightforwardly, it is still worthy understanding this formula via the BS Feynman diagrams in a more explicit way: (i) On one hand, if we try to apply Jacobi identities to reduce the BCJ double-copy formula of GR integrand in terms of YM integrands, we have to beware of the loop momentum in the other copy of numerators. The loop momentum in distinct copies may be not aligned in this process. Thus the expected expansion formula for GR integrand (EYM integrand) is far from apparent from this aspect. (ii) On another hand, once EYM (and thus GR) expansion formula has been obtained, its connection to the local expressions of EYM (GR) integrands can be established, following from the algorithm proposed in \cite{Xie:2025utp}.

In this work, we provide a rigorous proof of the EYM expansion formula using explicit BS Feynman diagrams. We first express a one‑loop EYM integrand as a sum of tree‑level colour‑stripped YM amplitudes in the forward limit, then further expand these YM amplitudes into BS amplitudes. We demonstrate that when the coefficients in both steps satisfy the consistency conditions, the EYM amplitude can ultimately be written as a combination of one‑loop YM integrands with quadratic propagators, while retaining the same coefficients as those appearing in the YMS expansion formula. Combining this result with the known relation between GR and EYM integrands \cite{He:2016mzd,He:2017spx,Geyer:2017ela,Porkert:2022efy,Xie:2025utp,Cao:2025ygu} yields an expansion formula for GR integrands in terms of YM integrands.



The structure of this paper is given as follows. In \secref{sec:review}, we provide a brief review of the forward limit approach of YM and YMS integrands as well as the consistency conditions for expansion coefficients. The forward limit expressed EYM integrands, in terms of tree-level BS amplitudes, are proposed in \secref{sec:relation}. In \secref{sec:Example}, we introduce an illustrative example to show how the expansion coefficients in EYM expansion formula are inherited from the corresponding YMS expansion formula.
The general proof is presented in \secref{sec:Proof}. We summarise this work in \secref{sec:Conclusions}. Helpful formulas are included in the appendix.

\section{Forward limit, YM, YMS integrands, and consistency condition of numerators}\label{sec:review}

The forward limit approach states that an $n$-point one-loop integrand can be expressed in terms of $(n+2)$-point tree-level amplitudes where two particles $\pm$ carry the loop momentum  $\pm\ell^{\mu}$ in opposite directions. For single-trace colour-ordered Yang-Mills integrands, this can be carried out in two successive steps: (i) first expressing the tree-level $(n+2)$-point amplitudes $A^{\text{YM}}$ (with $\pm$ are fixed as the first and the last elements) in terms of the Kleiss-Kuijf (KK) \cite{Kleiss:1988ne} basis of BS ones $A^{\text{BS}}$ (where $\pm$ are fixed as the two ends) according to the tree-level expansion formula \cite{Fu:2017uzt}, and (ii) taking the forward limit $k^{\mu}_{\pm}\to \pm\ell^{\mu}$ of the tree amplitudes. The integrand ${\cal I}^{\text{YM}}(1,2,...,n)$ constructed in this way is explicitly presented as
\bea
{\cal I}^{\text{YM}}(1,2,...,n)&=&{1\over \ell^2}\lim\limits_{k_{\pm}\to \pm\ell}\Sl_{h}A^{\text{YM}}(+,1,2,...,n,-)+\text{cyc} (1,...,n)\nn
&=&{1\over \ell^2}\lim\limits_{k_{\pm}\to \pm\ell}\Sl_{h}\left[\,\Sl_{\pmb\sigma\in \text{S}_n}N(+,\pmb\sigma,-)A^{\text{BS}}(+,\pmb\sigma,-|+,1,2,...,n,-)\,\right]+\text{cyc} (1,...,n),\label{Eq:ForwardYM0}
\eea
where the $\text{cyc} (1,...,n)$ implies that all cyclic orderings of $1$, ..., $n$ should be summed over. In the above equation, we formally summed over gluon states $h$, while these are further reduced together with the polarisations $\epsilon^{\mu}_{\pm}$ as 
\bea
\Sl_{h}\epsilon_{-}^{\mu}\epsilon_{+}^{\nu}\to\Delta^{\mu\nu},~~\eta_{\mu\nu}\Delta^{\mu\nu}\to D-2,~~V_{\mu}W_{\nu}\Delta^{\mu\nu}\to V\cdot W. \label{Eq:Polarizations}
\eea
The $D$ refers to the dimension of spacetime in which the loop momentum $\ell^{\mu}$ lives, when dimensional regularisation is taken. The $V_{\mu}$ and $W_{\nu}$ are two arbitrary vectors.
In the squarebrackets on the second line,   all permutations $\pmb\sigma\in \text{S}_n$ of external particles are summed over. The $\text{cyc} (1,...,n)$ means we also sum over all terms related by cyclic permutations. The expansion coefficients $N(+,\pmb\sigma,-)$ in (\ref{Eq:ForwardYM0}) are functions of 
Lorentz contractions of loop momentum $\ell^{\mu}$, external momenta $k^{\mu}_i$ and external polarisarions $\epsilon^{\mu}_j$ of gluons. These functions form the Del Duca-Dixon-Maltoni (DDM) basis \cite{DelDuca:1999rs,Bern:2010yg} of BCJ numerators which generates all other tree-level BCJ numerators via algebraic identities.

 It will be convenient to fix the position of loop momentum $\ell^{\mu}$ in the integrand as the one to the left of gluon $1$. Consequently, the expression (\ref{Eq:ForwardYM0}) of YM integrand turns into 
\bea
{\cal I}^{\text{YM}}(\ell;1,2,...,n) &\equiv&\Sl_{h}\Biggl[\,\Sl_{\pmb\sigma\in \text{S}_n}{1\over (\ell-k_{\pmb\sigma_L})^2}N(+(\ell-k_{\pmb\sigma_L}),\pmb\sigma,-(\ell-k_{\pmb\sigma_L}))\\
&&\times 
A^{\text{BS}}(+(\ell-k_{\pmb\sigma_L}),\pmb\sigma,-(\ell-k_{\pmb\sigma_L})|+(\ell-k_{\pmb\sigma_L}),1,2,...,n,-(\ell-k_{\pmb\sigma_L}))\Biggr]+\text{cyc} (1,...,n).
\label{Eq:ForwardYM}
\eea
Suppose each $\pmb\sigma$ is written as $(\pmb\sigma_L,1,\pmb\sigma_R)$, where $\pmb\sigma_L,$ and $\pmb\sigma_R$ denote the two ordered subsets separated by 1. Comparing to (\ref{Eq:ForwardYM0}), the loop momentum in each term of the above expression has been shifted as $\ell^{\mu}\to \ell^{\mu}-k^{\mu}_{\pmb\sigma_L}$ so that $\ell^{\mu}$ is fixed as the one to the left of $1$. Note that this overall shift only yields terms that integrate to zero in dimensional regularisation, and thus does not affect the loop integral.

The doubly colour-ordered scalar-loop YMS integrand ${\cal I}^{\text{YMS}}(\ell;1,2,...,r||\mathsf{G}\,|\,\pmb\rho)$  with scalars $1$, $2$, ..., $r$ coupled to gluons in $\mathsf{G}$ also obeys a similar expression in terms of BS ones, based on the forward limit of the corresponding tree-level expansion formula \cite{Fu:2017uzt}
\bea
{\cal I}^{\text{YMS}}(\ell;1,2,...,r||\mathsf{G}\,|\,\pmb\rho)&\equiv&
%
%
%
%
\Biggl\{\bigg[\,\Sl_{\pmb\sigma\in \{1,...,r\}\shuffle\text{perms}\,\mathsf{G}}{1\over (\ell-k_{\pmb\sigma_L})^2}C(+(\ell-k_{\pmb\sigma_L}),\pmb\sigma,-(\ell-k_{\pmb\sigma_L}))\nn
&&~~~~~~~~~~~\times A^{\text{BS}}(+(\ell-k_{\pmb\sigma_L}),\pmb\sigma,-(\ell-k_{\pmb\sigma_L})||\mathsf{G}\,|\,+(\ell-k_{\pmb\sigma_L}),\pmb\rho,-(\ell-k_{\pmb\sigma_L}))\,\bigg]\nn
&&~~~~~~~~~~~~~~~~~~~~~~~~~~~~~~~~~~~~~~~~~~~~~~~~~~+\text{cyc} (1,...,r)\Biggr\}+\text{cyc}(\pmb\rho),\label{Eq:ForwardYMs}
\eea
in which $C(+(\ell-k_{\pmb\sigma_L}),\pmb\sigma,-(\ell-k_{\pmb\sigma_L}))$ denotes the kinematic coefficients which are functions of Lorentz contractions of $\ell^{\mu}$, polarsations and external momenta. On the second line, we introduced the shuffle $A\shuffle B$ of two ordered sets $A$ and $B$, which is defined by the permutations of all elements of the two sets so that the relative ordering in each ordered set is perserved. For example, $\{1,2\}\shuffle\{3,4\}$ is the collection of the following permutations
\bea
\{1,2,3,4\},~~\{1,3,2,4\},~~\{1,3,4,2\},~~\{3,1,2,4\},~~\{3,1,4,2\},~~\{3,4,1,2\}.
\eea
The $\text{perms}\,\mathsf{G}$ denotes all possible permutations of gluons in $\mathsf{G}$.
In (\ref{Eq:ForwardYMs}), we have summed over all terms related by cyclic permutations of scalars $1$, ..., $r$, and the cyclic permutations of  $\pmb\rho\in \text{perms}\,(\{1,...,r\}\cup\mathsf{G})$. As we have done in (\ref{Eq:ForwardYM}), we have fixed the loop momentum  $\ell^{\mu}$ as the one attached to the scalar $1$ from left.

Though the expansion formulas (\ref{Eq:ForwardYM}) and (\ref{Eq:ForwardYMs}) generate the one-loop integrands of YM and YMS, these expressions are based on the forward limit of tree-level BS amplitudes, in which the loop propagators are linear propagators which have the form
\bea
 {1\over 2 \ell\cdot k_A+k_A^2}, \label{Eq:LinearPropagator}
 \eea
where $k_A^{\mu}$ denotes the total momenta of external particles in the set $A$. This expression is apparently different from the quadratic propagator ${1\over 2 (\ell+k_A)^2}$ in conventional Feynman diagrams.
Generally speaking, the expansion formulas  (\ref{Eq:ForwardYM}) and (\ref{Eq:ForwardYMs}) cannot be directly extended to expansion formulas of YM, YMS integrands in terms of conventional loop BS integrands with quadratic propagators. In addition, despite the coefficients in (\ref{Eq:ForwardYM}) and (\ref{Eq:ForwardYMs}) reflect the tree-level BCJ numerators in the DDM basis, they cannot be naively considered as the one-loop numerators.

As revealed in \cite{Du:2025yxz}, in order to construct correct BCJ numerators, we can impose additional consistency conditions on the coefficients in (\ref{Eq:ForwardYM}) and (\ref{Eq:ForwardYMs}) 
\bea
N(\ell,\sigma_1=1,\sigma_2,...,\sigma_n,-\ell)&=&N(\ell+k_1,\sigma_2,...,\sigma_n,1,-(\ell+k_1))\nn
&=&...=N(\ell+k_{1,\sigma_{n-1}},\sigma_n,1,\sigma_2,...,\sigma_{n-1},-(\ell+k_{1,\sigma_{n-1}})),\label{Eq:ConsistencyYM}\\
C(\ell,\sigma_1=1,\sigma_2,...,\sigma_n,-\ell)&=&C(\ell+k_1,\sigma_2,...,\sigma_n,1,-(\ell+k_1))\nn
&=&...=C(\ell+k_{1,\sigma_{n-1}},\sigma_n,1,\sigma_2,...,\sigma_{n-1},-(\ell+k_{1,\sigma_{n-1}})),\label{Eq:ConsistencyYMs}
\eea
where $k_{1,\sigma_{n-1}}\equiv k_1+k_{\sigma_2}+\dots+k_{\sigma_{n-1}}$. The $\pm$ particles are explicitly denoted by their momenta. Once the consistency conditions are satisfied, they can play as the consistency coefficients when expressing YM or YMS integrands in terms of the propagator matrix ${\cal I}^{\text{BS}}$ (i.e., the BS integrands with quadratic propagators) 
\bea
{\cal I}^{\text{YM}}(\ell;1,2,...,n)&=&\Sl_{\{\sigma_2,...,\sigma_n\}\in\text{S}_{n-1}}N(\ell;1,\sigma_2,...,\sigma_n)\,{\cal I}^{\text{BS}}(\ell;1,\sigma_2,...,\sigma_n|1,2,...,n),\label{Eq:ExpYM}\\
{\cal I}^{\text{YMS}}(\ell;1,2,...,r||\mathsf{G}\,|\,\pmb\rho)&=&\Sl_{\{\sigma_2,...,\sigma_n\}\in\{2,...,r\}\shuffle\text{perms}\,\mathsf{G}}C(\ell;1,\sigma_2,...,\sigma_n)\,{\cal I}^{\text{BS}}(\ell;1,\sigma_2,...,\sigma_n\,|\,\pmb\rho),\label{Eq:ExpYMs}
\eea
in which the kinematic coefficients $N(\ell;1,\sigma_2,...,\sigma_n)$ and $C(\ell;1,\sigma_2,...,\sigma_n)$ are defined by those satisfying the consistency conditions \eqref{Eq:ConsistencyYM} and \eqref{Eq:ConsistencyYMs} in (\ref{Eq:ExpYM})
\bea
N(\ell;1,\sigma_2,...,\sigma_n)\equiv N(\ell,\sigma_1=1,\sigma_2,...,\sigma_n,-\ell),~~C(\ell;1,\sigma_2,...,\sigma_n)=C(\ell,\sigma_1=1,\sigma_2,...,\sigma_n,-\ell).
\eea
These kinematic coefficients are now the proper ones for constructing BCJ numerators. 

For the YMS expansion formula (\ref{Eq:ExpYMs}), explicit coefficients for integrands with up to three gluons were proposed \cite{Du:2025yxz}
\bea
C(\ell;1,\pmb\sigma)=\epsilon_p\cdot X_p(\ell;1,\pmb\sigma)\,,\label{Eq:CoefficientEG1a}
\eea
\bea
C(\ell;1,\pmb\sigma)=\left[(\epsilon_p)_{\mu}(\epsilon_q)_{\nu}-{\epsilon_p\cdot\epsilon_q\over k_p\cdot k_q}\,(k_p)_{\mu}(k_q)_{\nu}\right]X^{\mu}_p(\ell;1,\pmb\sigma)X^{\mu}_q(\ell;1,\pmb\sigma)\,,\label{Eq:YMSExp2GravitonsCF}
\eea

\bea
\begin{matrix}
C(\ell;1,\pmb\sigma)&=&\biggl[\,(\epsilon_p)_{\mu}(\epsilon_q)_{\nu}(\epsilon_s)_{\tau}-{\epsilon_p\cdot\epsilon_q\over k_p\cdot k_q}\,(k_p)_{\mu}(k_q)_{\nu}(\epsilon_s)_{\tau}&\\
&&-{\epsilon_p\cdot\epsilon_s\over k_p\cdot k_s}\,(k_p)_{\mu}(\epsilon_q)_{\nu}(k_s)_{\tau}-{\epsilon_q\cdot\epsilon_s\over k_q\cdot k_s}\,(\epsilon_p)_{\mu}(k_q)_{\nu}(k_s)_{\tau}&\biggr]X^{\mu}_p(\ell;1,\pmb\sigma)X^{\mu}_q(\ell;1,\pmb\sigma)X^{\tau}_{s}(\ell;1,\pmb\sigma)
\end{matrix},\label{Eq:YMSExp3GravitonsCF}
\eea
where $p$, $q$, $s$ denotes the gluons, while $X^{\mu}_{i}(\ell;\pmb\sigma)$ $(i=p,q, \text{or}, s)$ refers to the sum of $\ell^{\mu}$ and the total momentum of external particles between  $\ell^{\mu}$ and the gluon $i$.

In the coming sections, we show the expansion formula for EYM integrand (which involves gravitons, dilatons and B-fields) parallel with the YMS one (\ref{Eq:ExpYMs}), under the assumption that the consistency numerators $N(\ell;1,\sigma_2,...,\sigma_n)$ and $C(\ell;1,\sigma_2,...,\sigma_n)$ in (\ref{Eq:ExpYM}) and  (\ref{Eq:ExpYMs}) are already available.

\section{The relation between EYM and YM integrands}\label{sec:relation}

According to the worldsheet approach \cite{He:2016mzd,He:2017spx,Geyer:2017ela,Porkert:2022efy}, the colour-ordered EYM integrand $\mathcal{I}^{\,\text{EYM}}(\ell;1...,r||\mathsf{H})$ where $1$,..., $r$ denote gluons and $\mathsf{H}$ denotes the graviton set, like the YM and YMS ones, also arises from the forward limit of the tree-level expansion formula 
\bea
\mathcal{I}^{\,\text{EYM}}(\ell;1...,r||\mathsf{H})&=&\Sl_{\pmb\sigma\in\{1,...,r\}\shuffle \text{perms}\,\mathsf{H} }{1\over (\ell-k_{\pmb\sigma_L})^2}C(\ell-k_{\pmb\sigma_L},\pmb\sigma,-(\ell-k_{\pmb\sigma_L}))\label{eq:gravity-half-ladder-welded0}\\
&&~~~~~~~~~~~~~~~~~~~~~~~~~~~\times A^{\text{YM}}(+(\ell-k_{\pmb\sigma_L}),\pmb\sigma_L,1,\pmb\sigma_R,-(\ell-k_{\pmb\sigma_L}))+\text{cyc}(1,...,r),\nonumber
\eea
 where we have applied the fact that tree-level EYM amplitudes can be expanded in terms of colour-ordered YM ones, sharing the same expansion coefficients as those for YMS amplitudes \cite{Fu:2017uzt}. The summation over $h$ in this formula and in the coming discussions are absorbed for brevity. For a given permutation $\pmb\sigma$, we again use $\pmb\sigma_L$ and  $\pmb\sigma_R$ to denote the ordered subsets separated by the gluon $1$, i.e., $\pmb\sigma=\pmb\sigma_L,1,\pmb\sigma_R$.
When the tree-level amplitudes $A^{\text{YM}}(+(\ell-k_{\pmb\sigma_L}),\pmb\sigma_L,1,\pmb\sigma_R,-(\ell-k_{\pmb\sigma_L}))$ are further expanded in terms of BS ones, the integrand (\ref{eq:gravity-half-ladder-welded0}) turns into 
\bea
\mathcal{I}^{\,\text{EYM}}(\ell;1...,r||\mathsf{H})
&=&\Sl_{\pmb\sigma\in\{1,...,r\}\shuffle \text{perms}~\mathsf{H} }{1\over (\ell-k_{\pmb\sigma_L})^2}C(\ell-k_{\pmb\sigma_L},\pmb\sigma,-(\ell-k_{\pmb\sigma_L}))\nn
&&~~~\times\Sl_{\pmb\rho\in\text{perms}~(\{1,...,r\}\cup\mathsf{H}) }~ A^{\text{BS}}(\ell-k_{\pmb\sigma_L},\pmb\sigma,-(\ell-k_{\pmb\sigma_L})|\ell-k_{\pmb\sigma_L},\pmb\rho,-(\ell-k_{\pmb\sigma_L}))\nn
&&~~~~~~~\times\W N(\ell-k_{\pmb\sigma_L},\pmb\rho,-(\ell-k_{\pmb\sigma_L}))
+\text{cyc}(1,...,r).\label{EYMeg0}
\eea
In the above expression, we have introduced the notation $\W N$ instead of  $N$ in (\ref{Eq:ForwardYM}), to emphasise that the polarisations in $\W N$ are different from those  in $C$. Specifically, $\W N$ involves polarisation vectors of gluons and one half polarisation tensors of gravitons, while $C$ only gets contribution from the other half polarisation tensors of gravitons. It is worthy  pointing out that in the right permutation  of a BS amplitude $A^{\text{BS}}(\ell-k_{\pmb\sigma_L},\pmb\sigma,-(\ell-k_{\pmb\sigma_L})|\ell-k_{\pmb\sigma_L},\pmb\rho,-(\ell-k_{\pmb\sigma_L}))$ the first and the last elements still carry momenta $\pm(\ell-k_{\pmb\sigma_L})$ which rely on the left permutation $\pmb\sigma=(\pmb\sigma_L,1,\pmb\sigma_R)$, respectively.

 In the next section, we prove that if all coefficients $C(+,\pmb\sigma,-)$ and  $\W N(+,\pmb\rho,-)$ satisfy the consistency conditions (\ref{Eq:ConsistencyYM}) and (\ref{Eq:ConsistencyYMs}), the EYM integrand can be expressed in terms of YM integrands with the same coefficients  $C(\ell;1,\sigma_2,...,\sigma_n)$ as those appearing in (\ref{Eq:ExpYMs})
\bea
\boxed{{\cal I}^{\text{EYM}}(\ell;1,2,...,r||\mathsf{H}\,)=\Sl_{\{\sigma_2,...,\sigma_n\}\in\{2,...,r\}\shuffle\text{perms}\,\mathsf{H}}C(\ell;1,\sigma_2,...,\sigma_n)\,{\cal I}^{\text{YM}}(\ell;1,\sigma_2,...,\sigma_n)}.\label{Eq:ExpEYM}
\eea
For integrands containing no more than three gravitons, the coefficients $C(\ell;1,\sigma_2,...,\sigma_n)$ are given explicitly by the compact expressions (\ref{Eq:CoefficientEG1a}), (\ref{Eq:YMSExp2GravitonsCF}), and (\ref{Eq:YMSExp3GravitonsCF}).

\section{An example study: the EYM integrand with two gluons and two gravitons }\label{sec:Example}

In this section, we examine the EYM integrand $\mathcal{I}^{\text{EYM}}(\ell;1,2|{p,q})$ with two gluons $1$, $2$ and two gravitons $p$, $q$ as a simple illustrative example. Our goal is to demonstrate how the forward limit representation (\ref{EYMeg0}), which contains linear propagators, can be reorganised into the expansion formula (\ref{Eq:ExpEYM}) that involves quadratic propagators, assuming that consistent coefficients are already supplied. The procedure consists of three steps: (i) First, we express the integrand in terms of tree‑level BS Feynman diagrams evaluated in the forward limit. (ii) Next, we convert the terms related by cyclic permutations into one-loop Feynman diagrams that are expressed with quadratic propagators. (iii) Finally, for a fixed right permutation we collect all corresponding contributions together. The consistency condition then allows this collection to be rewritten as a YM integrand. Consequently, we show that $\mathcal{I}^{\text{EYM}}(\ell;1,2|{p,q})$ indeed satisfies the expansion formula (\ref{Eq:ExpEYM}).

\subsection{Expressing the integrand in terms of tree-level BS Feynman diagrams }

Following (\ref{EYMeg0}), the EYM integrand $\mathcal{I}^{\text{EYM}}(\ell;1,2|\{p,q\})$  is explicitly given by
\bea
\mathcal{I}^{\,\text{EYM}}(\ell;1,2||\{p,q\})
&=&\Sl_{\pmb\sigma\in\{1,2\}\shuffle \text{perms}\{p,q\} }{1\over (\ell-k_{\pmb\sigma_L})^2}C(\ell-k_{\pmb\sigma_L},\pmb\sigma,-(\ell-k_{\pmb\sigma_L}))\nn
&&~~~\times\Sl_{\pmb\rho\in\text{perms}~(\{1,2\}\cup\{p,q\}) }~ A^{\text{BS}}(\ell-k_{\pmb\sigma_L},\pmb\sigma,-(\ell-k_{\pmb\sigma_L})|\ell-k_{\pmb\sigma_L},\pmb\rho,-(\ell-k_{\pmb\sigma_L}))\nn
&&~~~~~~~\times\W N(\ell-k_{\pmb\sigma_L},\pmb\rho,-(\ell-k_{\pmb\sigma_L}))
+\text{cyc}(1,2).\label{EYMExpEG1}
\eea
Here the integrand has been decomposed into a combination of doubly colour-ordered tree-level BS amplitude. In this example there are 12 distinct left permutations $\pmb\sigma\in\{1,2\}\shuffle \text{perms}\{p,q\}$ 
\bea
\{1,2,p,q\},~~\{1,p,2,q\},~~\{1,p,q,2\},~~\{p,1,2,q\},~~\{p,1,q,2\},~~\{p,q,1,2\},~~(p\leftrightarrow q),
\eea
and $4!=24$ possible right permutations $\pmb\rho\in\text{perms}~(\{1,2\}\cup\{p,q\})$, i.e. all permutations of the four external particles. Including cyclic permutations of $(1,2)$ adds another 12 left permutations.

According to (\ref{Eq:BSForwardLimit}), a BS amplitude $A^{\text{BS}}(\ell-k_{\pmb\sigma_L},\pmb\sigma,-(\ell-k_{\pmb\sigma_L})|\ell-k_{\pmb\sigma_L},\pmb\rho,-(\ell-k_{\pmb\sigma_L}))$ for given $\pmb\rho$, $\pmb\sigma$ is expressed as the sum of all admissable Feynman diagrams, constructed by the following steps. 
\begin{itemize}

\item Divide the left and right permutations $\pmb\sigma\to \pmb\sigma_1|...|\pmb\sigma_I$ and $\pmb\rho\to \pmb\rho_1|...|\pmb\rho_I$ into the same number of ordered subsets such that each pair ($\pmb\sigma_i$, $\pmb\rho_i$) contains the same set of elements.

\item Associate a BS current $\phi_{\pmb\sigma_i|\pmb\rho_i}$ (defined in (\ref{Eq:BScurrent})) with each pair ($\pmb\sigma_i$, $\pmb\rho_i$). The corresponding Feynman diagram is then obtained by attaching these BS currents to the linear‑propagator line that connects the forward‑limit particles $\pm(\ell-k_{\pmb\sigma_L})$.

\end{itemize}
As a concrete example, take $\pmb\sigma=\{1,p,q,2\}\in \{1,\{2\}\shuffle\text{perms}\{p,q\}\}$ and $\pmb\rho=\{p,1,2,q\}\in\text{perms}\,(\{1,2\}\cup\{p,q\})$. The possible divisions are
\bea
\pmb\sigma\to \{1,p\}|\{q,2\},~~\pmb\rho\to \{p,1\}|\{2,q\};~~~\pmb\sigma\to \{1,p,q,2\},~~\pmb\rho\to \{p,1,2,q\}. \label{Eq:DivisionsEG1}
\eea
Hence, the BS amplitude is expressed as 
\bea
A^{\text{BS}}(\ell,1,p,q,2,-(\ell-k_{\pmb\sigma_L})|\ell-k_{\pmb\sigma_L},p,1,2,q,-\ell)&=&\begin{minipage}{1.8cm}\includegraphics[width=1.8cm]{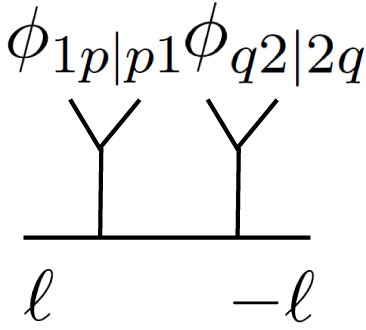}\end{minipage}+\begin{minipage}{1.6cm}\includegraphics[width=1.6cm]{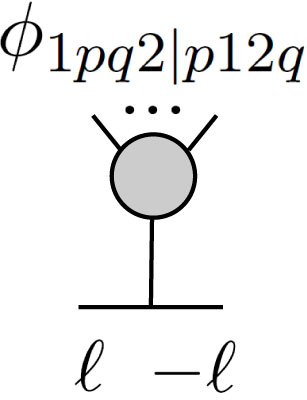}\end{minipage}\nn
&=&\phi_{1p|p1}\,{1\over 2\ell\cdot k_{1p}+k_{1p}^2}\,\phi_{q2|2q}+\phi_{1pq2|p12q},\label{Eq:BSEGdiagrams1}
\eea
where we have used the expression for the linear propagator (\ref{Eq:LinearPropagator}). Note that for this particular $\pmb\sigma$ we have $\pmb\sigma_L=\emptyset$, so $\pm(\ell-k_{\pmb\sigma_L})=\pm\ell$. All other BS amplitudes are expressed analogously in terms of tree-level Feynman diagrams.

Once every tree‑level BS amplitude is expressed as a sum of Feynman diagrams with linear propagators, the integrand becomes
\bea
\mathcal{I}^{\,\text{EYM}}(\ell;1,2||\{p,q\})
&=&\Sl_{\pmb\sigma\in\{1,2\}\shuffle \text{perms}\{p,q\} }{1\over (\ell-k_{\pmb\sigma_L})^2}C(\ell-k_{\pmb\sigma_L},\pmb\sigma,-(\ell-k_{\pmb\sigma_L}))\nn
&&\times\Sl_{\pmb\rho\in\text{perms}~(\{1,2\}\cup\{p,q\}) }\Sl_{\substack{\pmb\sigma\to \pmb\sigma_1|\pmb\sigma_2|...|\pmb\sigma_I\\ \pmb\rho\to \pmb\rho_1|\pmb\rho_2|...|\pmb\rho_I \\ |\pmb\sigma_i|=|\pmb\rho_i|}}\text{Diagram}\Bigl[\,\pmb\sigma\to \pmb\sigma_1|\pmb\sigma_2|...|\pmb\sigma_I\,\Big|\,\pmb\rho\to \pmb\rho_1|\pmb\rho_2|...|\pmb\rho_I \Bigr]\nn
&&~~~~~~~\times\W N(\ell-k_{\pmb\sigma_L},\pmb\rho,-(\ell-k_{\pmb\sigma_L}))
+\text{cyc}(1,2).\label{EYMExpEG2}
\eea
On the second line, we have formally written a BS amplitude for given $\pmb\sigma$ and $\pmb\rho$ as a sum of Feynman diagrams, each characterised by a pair of divisions $\pmb\sigma\to \pmb\sigma_1|\pmb\sigma_2|...|\pmb\sigma_I$ and $\pmb\rho\to \pmb\rho_1|\pmb\rho_2|...|\pmb\rho_I$. Note that 
for $\pmb\sigma=\{1,p,q,2\}$, $\pmb\rho=\{p,1,2,q\}$, the summation over divisions is precisely given by (\ref{Eq:BSEGdiagrams1}).

In the next subsection, we will focus on a pair of divisions together with those related by cyclic permutation of $(1,2)$, and 
demonstrate how the associated linear-propagator Feynman diagrams can be reorganised into diagrams with quadratic propagators.

\subsection{Converting linear propagators into quadratic ones for given pair of divisions}

We now study the two pairs of divisions (\ref{Eq:DivisionsEG1}) corresponding to $\pmb\sigma=\{1,p,q,2\}$, $\pmb\rho=\{p,1,2,q\}$. 

{\bf For the divisions $\pmb\sigma\to \{1,p\}|\{q,2\}$, $\pmb\rho\to \{p,1\}|\{2,q\}$}, we associate the coefficients  $C(\ell,1,p,q,2,-\ell)$ and  $\W N(\ell,p,1,2,q,-\ell)$ with the Feynman diagram and get
\bea
C(\ell,1,p,q,2,-\ell)\begin{minipage}{1.8cm}\includegraphics[width=1.8cm]{EYM1}\end{minipage}\W N(\ell,p,1,2,q,-\ell).\label{Eq:EYMeg1}
\eea
A further observation is that the summations over $\pmb\sigma$, $\pmb\rho$ in (\ref{EYMExpEG2}) and over the cyclic permutation of ($1,2$) together permit the following term
\bea
C(\ell-k_q-k_2,q,2,1,p,-(\ell-k_q-k_2))\begin{minipage}{4.4cm}\includegraphics[width=4.4cm]{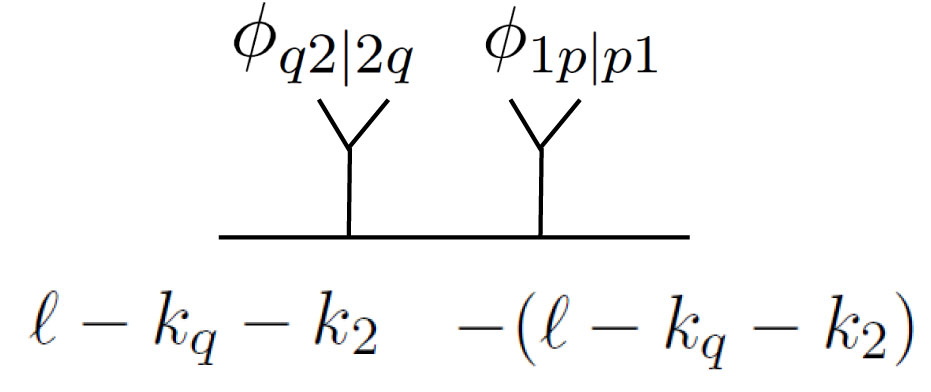}\end{minipage}\W N(\ell-k_q-k_2,2,q,p,1,-(\ell-k_q-k_2)),\label{Eq:EYMeg2}\nn
\eea
which comes from the BS amplitude with $\pmb\sigma=\{q,2,1,p\}$, $\pmb\rho=\{2,q,p,1\}$.
The crucial point is the following. Since the coefficients are supposed to satisfy the consistency conditions (\ref{Eq:ConsistencyYMs}) and (\ref{Eq:ConsistencyYM}) respectively, they can be reexpressed as 
\bea
C(\ell-k_q-k_2,q,2,1,p,-(\ell-k_q-k_2))&=&C(\ell,1,p,q,2,-\ell),\nn
\W N(\ell-k_q-k_2,2,q,p,1,-(\ell-k_q-k_2))&=&\W N(\ell,p,1,2,q,-\ell),\label{Eq:EYMeg4}
\eea
where the equality on the second line follows from redefining the loop momentum in (\ref{Eq:ConsistencyYM}) via $\ell\to \ell'=\ell+k_p$.

We now multiply  (\ref{Eq:EYMeg1}) and (\ref{Eq:EYMeg2}) by $1\over \ell^2$ and  $1\over (\ell-k_q-k_2)^2$ respectively and then sum them together.  Considering the consistency conditions (\ref{Eq:EYMeg4}) and noting that the Feynman diagrams in each sum are related by cyclic permutations of the BS subcurrents $\phi_{1p|p1}$ and $\phi_{q2|2q}$, we get the following term
\bea
&&C(\ell,1,p,q,2,-\ell)\left[\,{1\over\ell^2}\begin{minipage}{1.8cm}\includegraphics[width=1.8cm]{EYM1}\end{minipage}+{1\over(\ell-k_q-k_2)^2}\begin{minipage}{4.2cm}\includegraphics[width=4.2cm]{EYM3}\end{minipage}\,\right]\W N(\ell,p,1,2,q,-\ell)\nn
&=& C(\ell;1,p,q,2)~\begin{minipage}{1.8cm}\includegraphics[width=1.8cm]{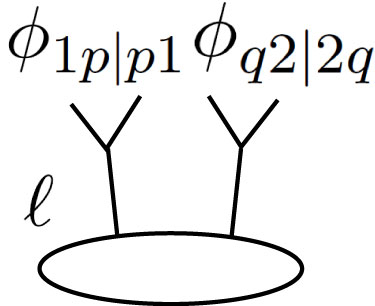}\end{minipage}~\W N(\ell;p,1,2,q)\nn
&=&C(\ell;1,p,q,2)~\left[\,{1\over\ell^2}\,\phi_{1p|p1}\,{1\over (\ell+k_{1p})^2}\,\phi_{q2|2q}\,\right]~\W N(\ell;p,1,2,q)
\label{Eq:EYMeg05}
\eea
where the coefficients are reexpressed by 
\bea
C(\ell;1,p,q,2)\equiv C(\ell,1,p,q,2,-\ell),~~~~\W N(\ell;p,1,2,q)\equiv\W N(\ell,p,1,2,q,-\ell).
\eea
In  (\ref{Eq:EYMeg05}), we have applied the identity (\ref{Eq:IDPartialFraction1}), through which the sum of linear-propagator  Feynman diagrams related by cyclic permutations of BS subcurrents is converted into a quadratic-propagator BS Feynman diagram.

{\bf For the divisions $\pmb\sigma\to \{1,p,q,2\}$, $\pmb\rho\to \{p,1,2,q\}$}, there is only a tadpole term in (\ref{EYMExpEG2}), whose contribution is shown as 
\bea
C(\ell;1,p,q,2)\,\left[\,{1\over \ell^2}\,\begin{minipage}{1.6cm}\includegraphics[width=1.6cm]{TadpoleEG1}\end{minipage}~\W N(\ell;p,1,2,q)\,\right]=C(\ell;1,p,q,2)\,\begin{minipage}{1.6cm}\includegraphics[width=1.6cm]{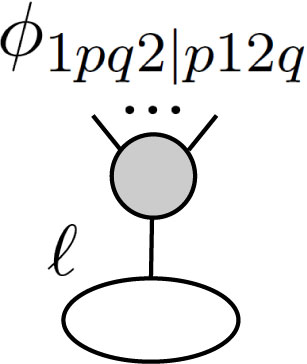}\end{minipage}~\W N(\ell;p,1,2,q),\label{Eq:EGTadpole1}
\eea
which is already expressed via quadratic propagator Feynman diagram. As demonstrated in \cite{Xie:2025utp}, the sum of all such tadpole diagrams that are related via cyclic permutations of gluons vanishes, due to $U(1)$-decoupling identity.

{\bf Contributions from cyclic permutations of $\pmb\sigma=\{q,2,1,p\}$, $\pmb\rho=\{2,q,p,1\}$}~~In the sums over  $\pmb\sigma\in\{1,2\}\shuffle \text{perms}\{p,q\}$ and $\pmb\rho\in\text{perms}~(\{1,2\}\cup\{p,q\})$ in (\ref{EYMExpEG2}), one can always find terms related to $\pmb\sigma=\{q,2,1,p\}$, $\pmb\rho=\{2,q,p,1\}$ via cyclic permutations. We repeat the above discussions on these terms, by collecting terms with linear-propagator Feynman diagrams into those with quadratic-propagator diagrams, for a given pair of $\pmb\sigma\in \text{cyc}\{q,2,1,p\}$ and $\pmb\rho\in\text{cyc}\{2,q,p,1\}$. In all diagrams we consistently choose the loop momentum $\ell^{\mu}$ as the one to the left of $1$ in the left permutations $\pmb\sigma\in \text{cyc}\{q,2,1,p\}$.

When considering all contributions of $\pmb\sigma=\{q,2,1,p\}$ and $\pmb\rho=\{2,q,p,1\}$ and those related by cyclic permutations, we arrive at
\bea
&&C(\ell;1,p,q,2)\,\left[\,\begin{minipage}{1.8cm}\includegraphics[width=1.8cm]{EYM6}\end{minipage}~\W N(\ell;p,1,2,q)+\begin{minipage}{2.6cm}\includegraphics[width=2.6cm]{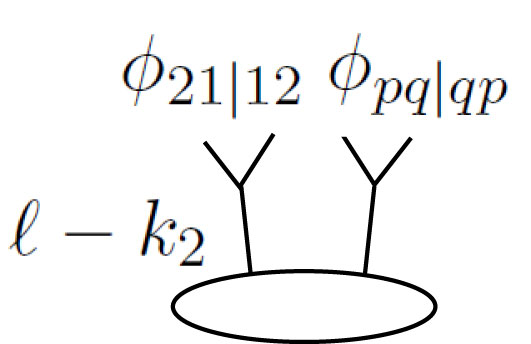}\end{minipage}~\W N(\ell-k_2;1,2,q,p)+\dots\right.\nn
&&~~~~~~~~~~~~~~~~~~\,\left.+\,\begin{minipage}{1.6cm}\includegraphics[width=1.6cm]{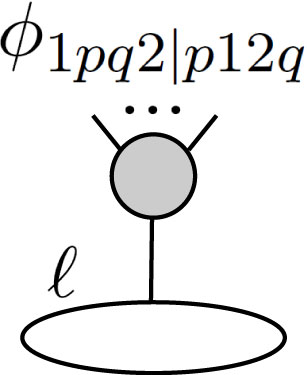}\end{minipage}~\W N(\ell;p,1,2,q)+\begin{minipage}{1.6cm}\includegraphics[width=1.6cm]{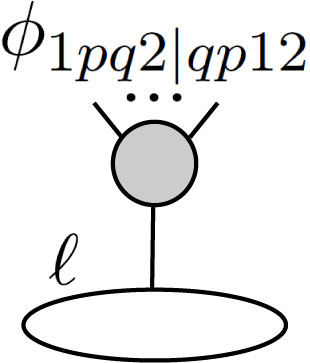}\end{minipage}~\W N(\ell;q,p,1,2)+\begin{minipage}{1.6cm}\includegraphics[width=1.6cm]{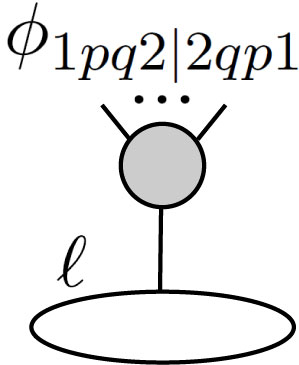}\end{minipage}~\W N(\ell;2,q,p,1)\right.\nn
&&~~~~~~~~~~~~~~~~~~\,\left.+\,\begin{minipage}{1.6cm}\includegraphics[width=1.6cm]{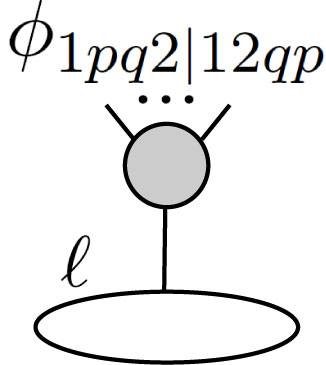}\end{minipage}~\W N(\ell;1,2,q,p)+\begin{minipage}{2.1cm}\includegraphics[width=2.1cm]{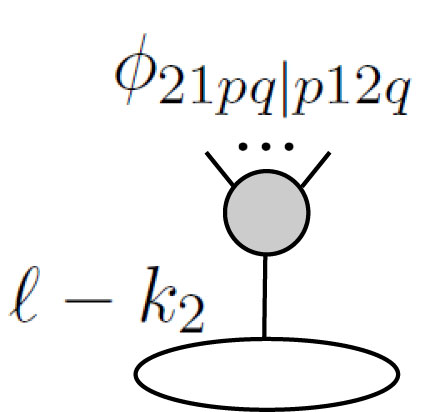}\end{minipage}~\W N(\ell-k_2;p,1,2,q)+\dots\,\right].\label{Eq:EYMeg06}
\eea
where we have applied the consistency condition (\ref{Eq:ConsistencyYMs}) for $C$, e.g. in the second term, we have considered
\bea
C(\ell-k_2;2,1,p,q)\begin{minipage}{2.2cm}\includegraphics[width=2.2cm]{EYM6a}\end{minipage}\W N(\ell-k_2;1,2,q,p)=C(\ell;1,p,q,2)\begin{minipage}{2.2cm}\includegraphics[width=2.2cm]{EYM6a}\end{minipage}\W N(\ell-k_2;1,2,q,p).\nn
\eea

\subsection{Expressing the EYM integrand in terms of YM integrands}

We now proceed to express the integrand $\mathcal{I}^{\,\text{EYM}}(\ell;1,2||\{p,q\})$ as a combination of YM integrands. To this end, we introduce notation  $\mathsf{T}(\ell)$ to stand for the expression inside the squarebrackets in (\ref{Eq:EYMeg06}):
\bea
\mathsf{T}(\ell)=\mathsf{T}^{(1)}+\mathsf{T}^{(2)}+\dots+\mathsf{T}^{(3)}+\mathsf{T}^{(4)}+\mathsf{T}^{(5)}+\mathsf{T}^{(6)}+\mathsf{T}^{(7)}+\dots,\label{Eq:EYMeg06a}
\eea
where the seven terms $\mathsf{T}^{(i)}$ correspond to the seven products (each a quadratic-propagator Feynman diagram multiplied by a right coefficient $\W N$) appearing within the brackets of (\ref{Eq:EYMeg06}). Suppose the subcurrent involving element $1$ is denoted as $\phi_{A,1,B|P,1,Q}$, with $A$, $B$ ($P$, $Q$) being the ordered sets separated by $1$ in the left (right) permutation. Applying an overall shift the loop momentum,  $\ell\to\ell-k_P+k_A$, to each term in $\mathsf{T}(\ell)$, we obtain
\bea
\mathsf{T}(\ell)\to \mathsf{T}'(\ell)&=&\mathsf{T}^{(1)}|_{\ell\to\ell-k_p}+\mathsf{T}^{(2)}|_{\ell\to\ell+k_2}+\dots+\mathsf{T}^{(3)}|_{\ell\to\ell-k_p}+\mathsf{T}^{(4)}|_{\ell\to\ell-k_p-k_q}\nn
&&+\mathsf{T}^{(5)}|_{\ell\to\ell+k_1}+\mathsf{T}^{(6)}+\mathsf{T}^{(7)}|_{\ell\to\ell+k_2-k_p}+\dots,
\eea
a transformation that leaves the loop integral invariant. Next, we apply the consistency condition (\ref{Eq:ConsistencyYM}) to the right coefficients, thereby rewriting all of them in terms of the standard coefficient $\W N(\ell;1,2,q,p)$, where the loop momentum is defined as the one to the left of particle 1 in the {\it right permutation}:
\bea
\mathsf{T}'(\ell)&=&\left[\,\begin{minipage}{2.3cm}\includegraphics[width=2.3cm]{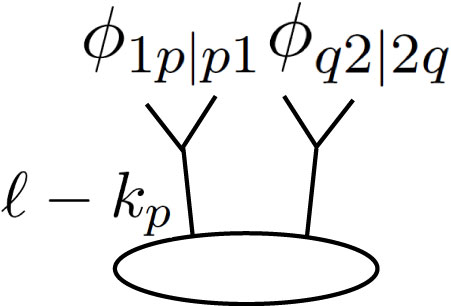}\end{minipage}~\W N(\ell-k_p;p,1,2,q)+\begin{minipage}{2.1cm}\includegraphics[width=2.1cm]{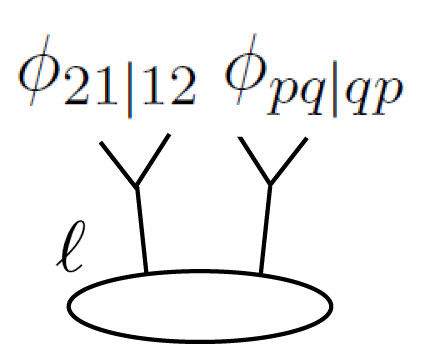}\end{minipage}~\W N(\ell;1,2,q,p)+\dots\right.\nn
&&~~\,\left.+\,\begin{minipage}{1.6cm}\includegraphics[width=1.6cm]{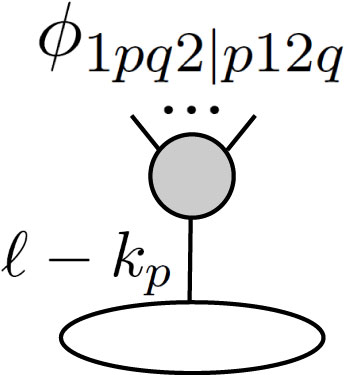}\end{minipage}~\W N(\ell-k_p;p,1,2,q)+\begin{minipage}{2.4cm}\includegraphics[width=2.3cm]{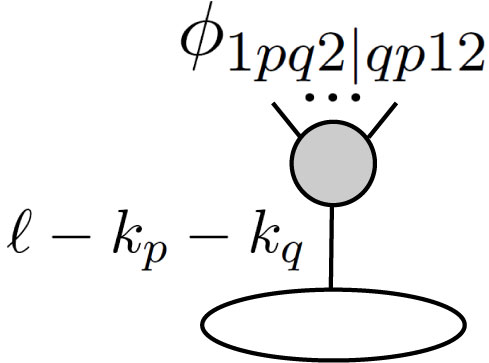}\end{minipage}~\W N(\ell-k_q-k_p;q,p,1,2)\right.\nn
&&~~\,\left.+\begin{minipage}{1.6cm}\includegraphics[width=1.6cm]{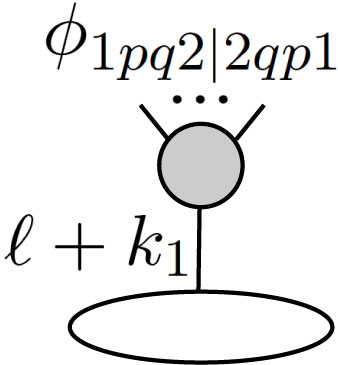}\end{minipage}\W N(\ell+k_1;2,q,p,1)+\begin{minipage}{1.5cm}\includegraphics[width=1.5cm]{TadpoleEG5}\end{minipage}\W N(\ell;1,2,q,p)+\begin{minipage}{1.7cm}\includegraphics[width=1.7cm]{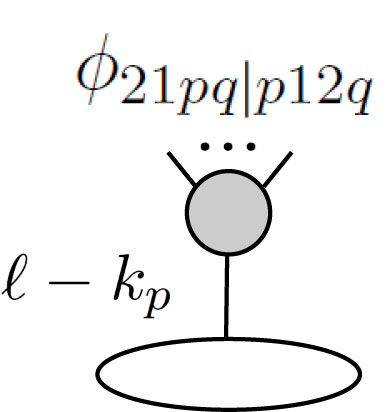}\end{minipage}\W N(\ell-k_p;p,1,2,q)+\dots\,\right]\nn
&=&\left[\,\begin{minipage}{2.3cm}\includegraphics[width=2.3cm]{EYM8}\end{minipage}+\begin{minipage}{2.1cm}\includegraphics[width=2.1cm]{TadpoleEG3a}\end{minipage}+\dots+~~\,\begin{minipage}{1.6cm}\includegraphics[width=1.6cm]{TadpoleEG6}\end{minipage}~~+~~\begin{minipage}{2.4cm}\includegraphics[width=2.3cm]{TadpoleEG7}\end{minipage}\right.\nn
&&~~~~~~~~~~~~~~~~~~~~~~~~\,\left.+~~\begin{minipage}{1.6cm}\includegraphics[width=1.6cm]{TadpoleEG8}\end{minipage}~~+~~\begin{minipage}{1.5cm}\includegraphics[width=1.5cm]{TadpoleEG5}\end{minipage}~~+~~\begin{minipage}{1.7cm}\includegraphics[width=1.7cm]{TadpoleEG4a}\end{minipage}~~+\dots\,\right]\W N(\ell;1,2,q,p)\nn
&=&{\cal I}^{\text{BS}}(1,p,q,2|\ell;1,2,q,p)\W N(\ell;1,2,q,p),\label{Eq:EYMeg06b}
\eea
where we have used the definition of BS integrand (\ref{Eq:BSQuadraticPropagatorIntegrand}).
This is nothing but just one term of the expansion formula (\ref{Eq:ExpYM}) of YM (with exchanging the roles of left and right and defining the position of loop momentum via the $1$ in the {\it right permutation}). 
Therefore, the $\mathsf{T}(\ell)$ in (\ref{Eq:EYMeg06a}) is also a term of the YM integrand, but the loop momentum is defined via the position of $1$ in the {\it left permutation!}

Guided by this observation, we rewrite the expression in (\ref{Eq:EYMeg06}) as 
\bea
&&C(\ell;1,p,q,2)\Big[T(\ell;1p|p1,q2|2q)~\W N(\ell;p,1,2,q)+T(\ell-k_2;21|12,pq|qp)~\W N(\ell-k_2;1,2,q,p)+\dots\nn
&&+T(\ell;1pq2|p12q)~\W N(\ell;p,1,2,q)+T(\ell;1pq2|qp12)~\W N(\ell;q,p,1,2)+T(\ell;1pq2|2qp1)~\W N(\ell;2,q,p,1)\nn
&&+T(\ell;1pq2|12qp)~\W N(\ell;1,2,q,p)+T(\ell-k_2;21pq|p12q)~\W N(\ell-k_2;p,1,2,q)+\dots\Big]\nn
&=&C(\ell;1,p,q,2)\left[\,\Sl_{T}T(\ell-k_A;A1B|P1Q,\dots)~\W N(l-k_A+k_P;1,2,q,p)\,\right],
\eea
where the sum runs over all possible Feynman diagrams $T(l-k_A;A1B|P1Q,\dots)$   (which  implies the subcurrents attached to the loop are $\phi_{A1B|P1Q}$,...) with respect to the permutations $\pmb\sigma=\{1,p,q,2\}$ and $\pmb\rho=\{p,1,2,q\}$ up to cyclic equivalence. 
Here, the loop momentum $\ell$ is  anchored by particle $1$ in the {\it left permutation}, and subcurrents containing $1$ have the form $\phi_{A1B|P1Q}$. The expression in the square brackets is equivalent to $\mathsf{T}'(\ell)={\cal I}^{\text{BS}}(1,p,q,2|\ell;1,2,q,p)\W N(\ell;1,2,q,p)$.

Extending this logic, we can group the terms in (\ref{EYMExpEG2}) with respect to permutations that are related by cyclic permutations of given $\pmb\sigma$ and $\pmb\rho$
\bea
\pmb\sigma\in\Big\{\{1,2\}\shuffle\text{perms}\{p,q\},\text{cyc} (1,2)\Big\}&=&\Big\{\pmb{\alpha}=\{1,\{2\}\shuffle\text{perms}\{p,q\}\},\text{cyc}~\pmb{\alpha}\Big\},\nn
\pmb\rho\in\text{perms}\,(\{1,2\}\cup\{p,q\})&=&\Big\{\pmb\beta=\{1,\text{perms}\{2,p,q\},\text{cyc}~\pmb{\beta}\Big\},
\eea
and then recast this part into expressions with quadratic propagators. The complete form of this part is
\bea
C(\ell;1,\pmb\sigma_{2,4})\left[\,\Sl_{T}T(\ell-k_A;A1B|P1Q,\dots)~\W N(l-k_A+k_P;1,\pmb\rho_{2,4})\,\right],
\eea
where $\pmb\sigma$ and $\pmb\rho$ have the form $1,\pmb\sigma_{2,4}$ and $1,\pmb\rho_{2,4}$ respectively, when $1$ is fixed as the first element using the cyclicity. The content within the square brackets is equivalent to ${\cal I}^{\text{BS}}(1,\pmb\sigma_{2,4}|\ell;1,\pmb\rho_{2,4})\W N(\ell;1,\pmb\rho_{2,4})$ (with ${\cal I}^{\text{BS}}$ defined by (\ref{Eq:BSQuadraticPropagatorIntegrand}) but with left and right permutations swapped), up to loop momentum shift  $\ell\to\ell-k_P+k_A$ in each term.

Summing over all cyclically equivalent $\pmb\rho$ and $\pmb\sigma$, (\ref{EYMeg0}) transforms into
\bea
&&\mathcal{I}^{\,\text{EYM}}(\ell;1,2||\{p,q\})\label{EYMeg7}\nn
&=&\Sl_{\pmb\sigma_{2,4}\in\{2\}\shuffle\text{perms}\{p,q\}}\,C(\ell;1,\pmb\sigma_{2,4}){\left[\,\Sl_{\pmb\rho_{2,4}\in\text{perms}\{2,p,q\} }\Sl_{T}T(\ell-k_A;A1B|P1Q,\dots)\W N(l-k_A+k_P;1,\pmb\rho_{2,4})\right]}\nn
&=&\Sl_{\pmb\sigma_{2,4}\in\{2\}\shuffle\text{perms}\{p,q\}}\,C(\ell;1,\pmb\sigma_{2,4})~\mathcal{I}^{\text{YM}}(\ell;1,\pmb\sigma_{2,4}).
\eea
where the expression inside the square brackets on the second line is just the YM integrand $\mathcal{I}^{\text{YM}}(\ell;1,\pmb\sigma_{2,4})$ which is expressed as a combination of BS one-loop Feynman diagrams with choosing $\ell$ as the one to the left of $1$ in the {\it left permutations}. It can be related to the standard expansion formula from as follows
\bea
&&\Sl_{\pmb\rho_{2,4}\in\text{perms}\{2,p,q\} }\Sl_{T}\left[T(\ell-k_A;A1B|P1Q,\dots)\W N(l-k_A+k_P;1,\pmb\rho_{2,4})\right]\bigg |_{\ell\to\ell-k_P+k_A}\nn
&=&\Sl_{\pmb\rho_{2,4}\in\text{perms}\{2,p,q\} }{\cal I}^{\text{BS}}(1,\pmb\sigma_{2,4}|\ell;1,\pmb\rho_{2,4})\W N(\ell;1,\pmb\rho_{2,4}),\label{Eq:EGshift}
\eea
where the loop momentum $\ell$ is fixed to the left of $1$ in the {\it right permutation}.

\section{Proof of the expansion formula of EYM integrand}\label{sec:Proof}
In this section, we extend the discussion of the previous section to provide a general proof of \eqref{Eq:ExpEYM}, based on the forward limit expression (\ref{EYMeg0}) and the consistency conditions (\ref{Eq:ConsistencyYM}), (\ref{Eq:ConsistencyYMs}). We begin by expressing the forward limit of the tree-level BS amplitude $A^{\text{BS}}(\ell-k_{\pmb\sigma_L},\pmb\sigma,-(\ell-k_{\pmb\sigma_L})|\ell-k_{\pmb\sigma_L},\pmb\rho,-(\ell-k_{\pmb\sigma_L}))$ in (\ref{EYMeg0}) as a sum of all possible Feynman diagrams. This sum can be organised through the following steps.  
\begin{itemize}
\item Divide the left permutation $\pmb\sigma$ into ordered subsets $\pmb\sigma\to \pmb\sigma_1|\pmb\sigma_2|...|\pmb\sigma_I$. 

\item For a given division of  $\pmb\sigma$, divide the right permutation  $\pmb\rho\to \pmb\rho_1|\pmb\rho_2|...|\pmb\rho_I$ such that $\pmb\rho_i$ contains {\it exactly the same elements as $\pmb\sigma_i$}. 

\item Associate each pair of ordered subsets $\pmb\sigma_i$, $\pmb\rho_i$ with a BS subcurrent $\phi_{\pmb\sigma_i|\pmb\rho_i}$ attached to the linear propagator line between $+$ and $-$.
\end{itemize}
The tree-level BS amplitude $A^{\text{BS}}$ in the forward limit is then explicitly displayed as
\bea
&&A^{\text{BS}}(\ell-k_{\pmb\sigma_L},\pmb\sigma,-(\ell-k_{\pmb\sigma_L})|\ell-k_{\pmb\sigma_L},\pmb\rho,-(\ell-k_{\pmb\sigma_L}))\nn
&=&\Sl_{\substack{\pmb\sigma\to \pmb\sigma_1|\pmb\sigma_2|...|\pmb\sigma_I\\ \pmb\rho\to \pmb\rho_1|\pmb\rho_2|...|\pmb\rho_I }}\begin{minipage}{3.2cm}\includegraphics[width=3.2cm]{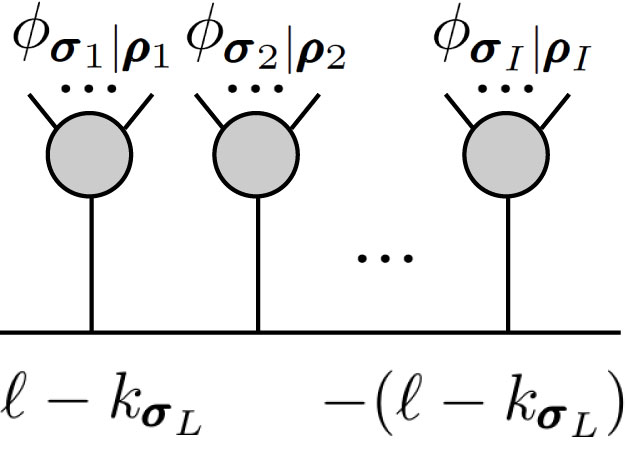}\end{minipage}\nn
&\equiv&\phi_{\pmb\sigma_1|\pmb\rho_1}\,{1\over 2k_{\pmb\sigma_1}\cdot \ell+k_{\pmb\sigma_1}^2}\,\phi_{\pmb\sigma_2|\pmb\rho_2}\,{1\over 2k_{\pmb\sigma_1,\pmb\sigma_2}\cdot \ell+k_{\pmb\sigma_1,\pmb\sigma_2}^2}\,\dots \,{1\over 2k_{\pmb\sigma_1,\pmb\sigma_{I-1}}\cdot \ell+k_{\pmb\sigma_1,\pmb\sigma_{I-1}}^2}\,\phi_{\pmb\sigma_I|\pmb\rho_I},
\label{Eq:GenFeynBSForward}
\eea
where $k_{\pmb\sigma_1,\pmb\sigma_i}\equiv k_{\pmb\sigma_1}+k_{\pmb\sigma_2}+...+k_{\pmb\sigma_i}$.

Substituting (\ref{Eq:GenFeynBSForward}) into (\ref{EYMeg0}), the EYM integrand $\mathcal{I}^{\,\text{EYM}}(\ell;1...,r||\mathsf{H})$ can be rewritten as 
\bea
\mathcal{I}^{\,\text{EYM}}(\ell;1...,r||\mathsf{H})&=&\Sl_{\substack{\pmb\sigma\in\{1,...,r\}\shuffle \text{perms}~\mathsf{H} \\ \pmb\rho\in\text{perms}~(\{1,...,r\}\cup\mathsf{H})}}\,\Sl_{\substack{\pmb\sigma\to \pmb\sigma_1|\pmb\sigma_2|...|\pmb\sigma_I\\ \pmb\rho\to \pmb\rho_1|\pmb\rho_2|...|\pmb\rho_I }}{1\over (\ell-k_{\pmb\sigma_L})^2}C(\ell-k_{\pmb\sigma_L},\pmb\sigma,-(\ell-k_{\pmb\sigma_L}))\label{EYMeg00}\nn
&&~~~~~~~\times \begin{minipage}{3.2cm}\includegraphics[width=3.2cm]{GenProofFeyn}\end{minipage}\W N(\ell-k_{\pmb\sigma_L},\pmb\rho,-(\ell-k_{\pmb\sigma_L}))+\text{cyc}(1,...,r).
\eea
On the rhs. of the above expression, for a given left permutation $\pmb\sigma\in\{1,...,r\}\shuffle \text{perms}~\mathsf{H}$, a given right permutation $\pmb\rho\in\text{perms}~(\{1,...,r\}\cup\mathsf{H})$ and a specific pair of divisions $\pmb\sigma\to \pmb\sigma_1|\pmb\sigma_2|...|\pmb\sigma_I$, $\pmb\rho\to \pmb\rho_1|\pmb\rho_2|...|\pmb\rho_I $, each pair of ordered subsets takes the general form
\bea
\pmb\sigma_i= \mathsf {S}_i\shuffle\text{perms}\,\mathsf{H}_i\equiv\left\{g^i_1,g^i_2,...,g^i_{j_i}\right\}\shuffle \text{perms}\,\left\{h^i_{1},...,h^{i}_{l_i}\right\},~~~~\pmb\rho_i=\left\{\rho^{i}_1,\rho^{i}_2,...,\rho^{i}_{j_i+l_i}\right\},
\eea
where $\big\{g^i_1,g^i_2,...,g^i_{j_i}\big\}$ denotes the ordered subset of gluons that respects the ordering $\{1,2,...,r\}$, $\big\{h^i_{1},...,h^{i}_{l_i}\big\}$ is the graviton subset, and $\big\{\rho^{i}_1,\rho^{i}_2,...,\rho^{i}_{j_i+l_i}\big\}$ stands for the ordered subset of gluons and gravitons that agrees with the full right permutation $\pmb\rho$. For a given pair of divisions in \eqref{EYMeg0}, we can always find other pairs of divisions in (\ref{EYMeg00}) that are related to the original pair via cyclic permutations of the ordered subsets:
\bea
&&\pmb\sigma_2|...|\pmb\sigma_I|\pmb\sigma_1,~~~~~~\,\,\pmb\rho_2|...|\pmb\rho_I|\pmb\rho_1,\nn
&&\pmb\sigma_3|...|\pmb\sigma_1|\pmb\sigma_2,~~~~~~\,\,\pmb\rho_3|...|\pmb\rho_1|\pmb\rho_2,\nn
&&\dots\nn
&&\pmb\sigma_I|\pmb\sigma_1|...|\pmb\sigma_{1-1},~~~~\pmb\rho_I|\pmb\rho_{1}|...|\pmb\rho_{i-1}.
\eea
Collecting these contributions together, we obtain 
\bea
&&C(\ell-k_{\pmb\sigma_1,\pmb\sigma_{jL}},\pmb\sigma_1,...,\pmb\sigma_I,-(\ell-k_{\pmb\sigma_1,\pmb\sigma_{jL}}))\begin{minipage}{3.5cm}\includegraphics[width=3.5cm]{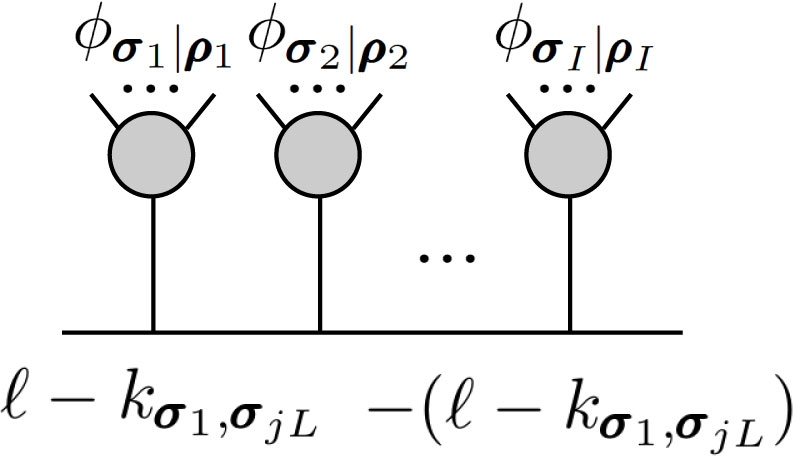}\end{minipage}{\W N(\ell-k_{\pmb\sigma_1,\pmb\sigma_{jL}},\pmb\rho_1,...,\pmb\rho_I,-(\ell-k_{\pmb\sigma_1,\pmb\sigma_{jL}}))\over (\ell-k_{\pmb\sigma_1,\pmb\sigma_{jL}})^2}\label{EYMeg1}\nn
 &+&C(\ell-k_{\pmb\sigma_2,\pmb\sigma_{jL}},\pmb\sigma_2,...,\pmb\sigma_1,-(\ell-k_{\pmb\sigma_2,\pmb\sigma_{jL}}))\begin{minipage}{3.5cm}\includegraphics[width=3.5cm]{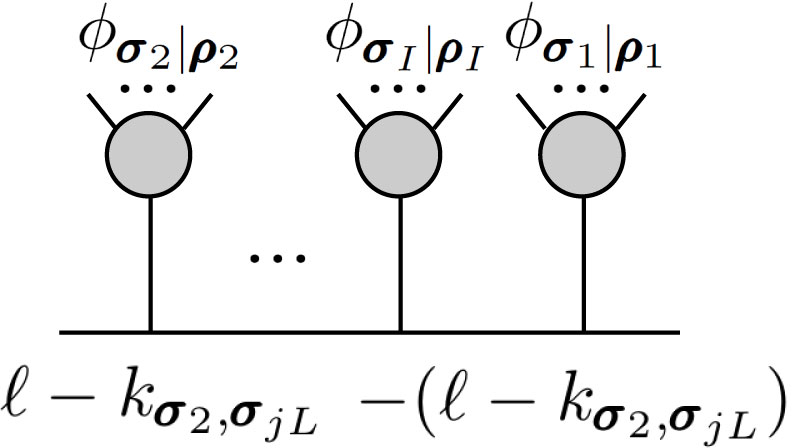}\end{minipage}{\W N(\ell-k_{\pmb\sigma_2,\pmb\sigma_{jL}},\pmb\rho_2,...,\pmb\rho_1,-(\ell-k_{\pmb\sigma_2,\pmb\sigma_{jL}}))\over (\ell-k_{\pmb\sigma_2,\pmb\sigma_{jL}})^2}\nn
 &+&...\nn
 &+&C(\ell-k_{\pmb\sigma_I,\pmb\sigma_{jL}},\pmb\sigma_I,...,\pmb\sigma_{I-1},-(\ell-k_{\pmb\sigma_I,\pmb\sigma_{jL}}))\begin{minipage}{3.3cm}\includegraphics[width=3.3cm]{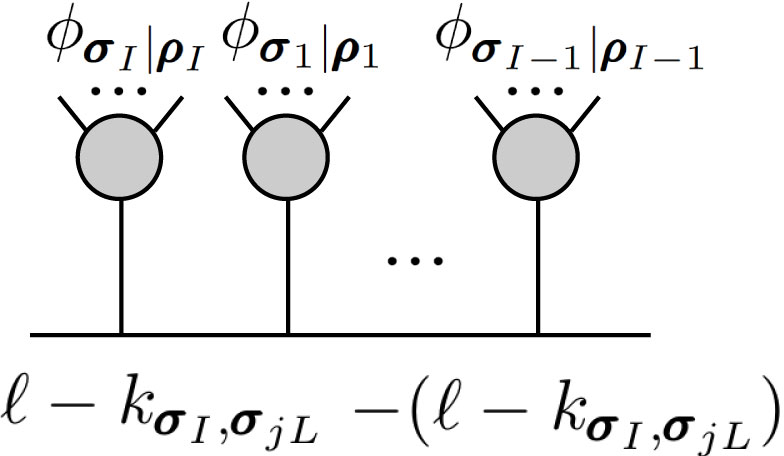}\end{minipage}{\W N(\ell-k_{\pmb\sigma_I,\pmb\sigma_{jL}},\pmb\rho_I,...,\pmb\rho_{I-1},-(\ell-k_{\pmb\sigma_I,\pmb\sigma_{jL}}))\over (\ell-k_{\pmb\sigma_I,\pmb\sigma_{jL}})^2}\nn
\eea
where $k_{\pmb\sigma_i,\pmb\sigma_{jL}}=k_{\pmb\sigma_i}+k_{\pmb\sigma_{i+1}}+\dots+k_{\pmb\sigma_{j-1}}+k_{\pmb\sigma_{jL}}$. Suppose the scalar $1$ is involved in $\pmb\sigma_j=\{\pmb\sigma_{jL},1,\pmb\sigma_{jR}\}$ and $\pmb\rho_j=\{\pmb\rho_{jL},1,\pmb\rho_{jR}\}$. The consistency conditions (\ref{Eq:ConsistencyYMs}), (\ref{Eq:ConsistencyYM})  allow us to express these coefficients as 
\bea
&&C(\ell-k_{\pmb\sigma_1,\pmb\sigma_{jL}},\pmb\sigma_1,...,\pmb\sigma_I,-(\ell-k_{\pmb\sigma_1,\pmb\sigma_{jL}}))=C(\ell-k_{\pmb\sigma_2,\pmb\sigma_{jL}},\pmb\sigma_2,...,\pmb\sigma_1,-(\ell-k_{\pmb\sigma_2,\pmb\sigma_{jL}}))=...\nn
&=&C(\ell-k_{\pmb\sigma_I,\pmb\sigma_{jL}},\pmb\sigma_I,...,\pmb\sigma_{I-1},-(\ell-k_{\pmb\sigma_I,\pmb\sigma_{jL}}))=C(\ell,1,\pmb\sigma_{jR},\pmb\sigma_{j+1},...,\pmb\sigma_{j-1},\pmb\sigma_{jL},-\ell),\\
&&\W N(\ell-k_{\pmb\sigma_1,\pmb\sigma_{jL}},\pmb\rho_1,...,\pmb\rho_I,-(\ell-k_{\pmb\sigma_1,\pmb\sigma_{jL}}))=\W N(\ell-k_{\pmb\sigma_2,\pmb\sigma_{jL}},\pmb\rho_2,...,\pmb\rho_1,-(\ell-k_{\pmb\sigma_2,\pmb\sigma_{jL}}))=...\nn
&=&\W N(\ell-k_{\pmb\sigma_I,\pmb\sigma_{jL}},\pmb\rho_I,...,\pmb\rho_{I-1},-(\ell-k_{\pmb\sigma_I,\pmb\sigma_{jL}}))=\W N(\ell-k_{\pmb\sigma_{jL}},\pmb\rho_j,...,\pmb\rho_{j-1},-(\ell-k_{\pmb\sigma_{jL}})).
\eea
Thus, the coefficients in each term of (\ref{EYMeg1}) can be factorised out as overall coefficients. According to the identity  (\ref{Eq:IDPartialFraction1}), the sum of diagrams in  (\ref{EYMeg1}) yields a  one-loop quadratic-propagator BS Feynman diagram. Therefore, (\ref{EYMeg1}) becomes
\bea
&&C(\ell;1,\pmb\sigma_{jR},\pmb\sigma_{j+1},...,\pmb\sigma_{j-1},\pmb\sigma_{jL})\begin{minipage}{3.9cm}\includegraphics[width=3.9cm]{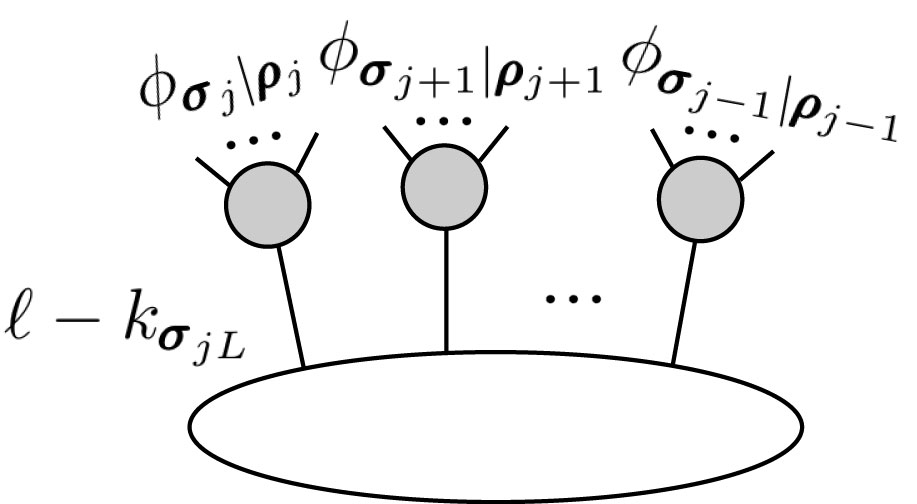}\end{minipage}\W N(\ell-k_{\pmb\sigma_{jL}};\pmb\rho_{j},\pmb\rho_{j+1},...,\pmb\rho_{j-1})\nn
&\equiv&C(\ell;1,\pmb\sigma_{jR},\pmb\sigma_{j+1},...,\pmb\sigma_{j-1},\pmb\sigma_{jL})\nn
&&~~~~~~~~~~~\times T\left(\ell-k_{\pmb\sigma_{jL}};\pmb\sigma_{j}|\pmb\rho_{j},\pmb\sigma_{j+1}|\pmb\rho_{j+1},\dots,\pmb\sigma_{j-1}|\pmb\rho_{j-1}\right)\times\,\W N(\ell-k_{\pmb\sigma_{jL}};\pmb\rho_{j},\pmb\rho_{j+1},...,\pmb\rho_{j-1}),
\eea
characterised by a pair of {\it cyclic divisions} (i.e., the cyclic equivalence has been quotiented out) of $\pmb\sigma$ and $\pmb\rho$.

Summing all the above terms corresponding to divisions of a given pair $\pmb\sigma\in\{1,...,r\}\shuffle \text{perms}~\mathsf{H}$ and  $\pmb\rho\in\text{perms}~(\{1,...,r\}\cup\mathsf{H})$ together, we obtain the sum of diagrams corresponding to all cyclic divisions of $\pmb\sigma$ and $\pmb\rho$, dressed by the coefficients $C$ and $\W N$:
\bea
C(\ell;1,\pmb\sigma_{jR},\pmb\sigma_{j+1},...,\pmb\sigma_{jL})\bigg[\Sl_{T}T(\ell-k_{\pmb\sigma_{jL}};\pmb\sigma_{j}|\pmb\rho_{j},\dots)~\W N(l-k_{\pmb\sigma_{jL}}+k_{\pmb\rho_{jL}};1,\pmb\rho_{jR},\pmb\rho_{j+1},...,\pmb\rho_{jL})\bigg],\label{EYMegNew1}
\eea
where we have further applied the consistency condition for $\W N$ so that $1$ is considered as the first element, recalling that $\pmb\rho_j=\{\pmb\rho_{jL},1,\pmb\rho_{jR}\}$. Similar to the example in the previous section, the expression inside square brackets becomes
\bea
&&\Sl_{T}\biggl[T(\ell-k_{\pmb\sigma_{jL}};\pmb\sigma_{j}|\pmb\rho_{j},\dots)~\W N(l-k_{\pmb\sigma_{jL}}+k_{\pmb\rho_{jL}};1,\pmb\rho_{jR},\pmb\rho_{j+1},...,\pmb\rho_{jL})\biggr]\bigg|_{\ell\to\ell-k_{\pmb\rho_{jL}}+k_{\pmb\sigma_{jL}}}\nn
&=&{\cal I}^{\text{BS}}(1,\pmb\sigma_{2,n}|\ell;1,\pmb\rho_{2,n})\W N(\ell;1,\pmb\rho_{2,n}),
\eea
with $\pmb\sigma$ and $\pmb\rho$ cyclically equivalent to $\{1,\pmb\sigma_{2,n}\}$, $\pmb\rho=\{1,\pmb\rho_{2,n}\}$, respectively.
Therefore (\ref{EYMegNew1}) corresponds to a single term of the YM integrand ${\cal I}^{\text{YM}}(\ell;1,\pmb\sigma_{jR},\pmb\sigma_{j+1},...,\pmb\sigma_{jL})$ when $\ell$ is fixed to the left of $1$ in the left permutation. 
Considering all permutations
\bea
&&\pmb\sigma\in\Big\{\{1,2,\dots,r\}\shuffle\text{perms}~\mathsf{H},\text{cyc} (1,\dots,r)\Big\}=\Big\{\pmb{\alpha}=\{1,\{2,\dots,r\}\shuffle\text{perms}~\mathsf{H}\},\text{cyc}~\pmb{\alpha}\Big\},\nn
&&\pmb\rho\in\text{perms}\,(\{1,2,...,r\}\cup\mathsf{H})=\Big\{\pmb\beta=\{1,\text{perms}\,(\{2,...,r\}\cup\mathsf{H})\},\text{cyc}~\pmb{\beta}\Big\}, 
\eea
we reexpress the integrand as 
\bea
\mathcal{I}^{\,\text{EYM}}(\ell;1...,r||\mathsf{H})&=&\Sl_{\pmb\sigma_{2,n}\in\{2,...,r\}\shuffle\text{perms}~\mathsf{H}}C(\ell;1,\pmb\sigma_{2,n})\nn
&&\times\Biggl[\,\Sl_{\pmb\rho_{2,n}\in\text{perms}(\{2,...,r\}\cup\mathsf{H})}\Sl_{T}T(\ell-k_{\pmb\sigma_{jL}};\pmb\sigma_{j}|\pmb\rho_{j},\dots)~\W N(l-k_{\pmb\sigma_{jL}}+k_{\pmb\rho_{jL}};1,\pmb\rho_{2,n})\,\Biggr]\nn
&=&\Sl_{\pmb\sigma_{2,n}\in\{2,...,r\}\shuffle\text{perms}~\mathsf{H}}C(\ell;1,\pmb\sigma_{2,n}){\cal I}^{\text{YM}}(\ell;1,\pmb\sigma_{2,n}).
\eea
Hence, we have completed the proof of (\ref{Eq:ExpEYM}).

\section{Conclusions}\label{sec:Conclusions}
In this work, we have provided a concrete proof of the statement that the gluon loop single-trace EYM integrands at one-loop level (with quadratic propagators) satisfy expansion relations, whereby an EYM integrand is expanded in terms of one-loop YM integrands expressed with quadratic propagators. The expansion coefficients in this formula share the same form as those in the corresponding expansion formula for scalar-loop YMS integrands. This proof is based on the forward limit representation of the EYM integrand and the consistency conditions for one-loop expansion coefficients. Since the existence of the consistent coefficients (for YM and YMS) has been proven in \cite{Du:2025yxz}, any expansion formula for YMS integrands with consistent coefficients implies the corresponding EYM expansion formula. Consequently, the explicit formula for YMS integrands \cite{Du:2025yxz} with up to three gluons directly yields the EYM expansion formula with up to three gravitions (with coefficients given by (\ref{Eq:CoefficientEG1a}), (\ref{Eq:YMSExp2GravitonsCF}) and (\ref{Eq:YMSExp3GravitonsCF})). This work may offer further insights for understanding the BCJ duality at the one-loop level.
\begin{acknowledgments}
We are grateful to Chih-Hao Fu and Yihong Wang for helpful discussions. YD is supported by NSFC under Grant No. 11875206.
\end{acknowledgments}

\appendix

\section{The Berends-Giele currents of BS theory}

The Berends-Giele current $\phi_{A|\W A}$, where $A$ and  $\widetilde{A}$ are ordered sets with the same number of elements, is defined as follows (\cite{Mafra:2016ltu}):
\bea
\phi_{A|\widetilde{A}}={1\over s_A}\Sl_{\substack{A\to A_L,A_R\\ \widetilde{A}\to\widetilde{A}_L,\widetilde{A}_R}}\Bigl[\,\phi_{A_L|\widetilde{A}_L}\phi_{A_R|\widetilde{A}_R}-\phi_{A_R|\widetilde{A}_L}\phi_{A_L|\widetilde{A}_R}\,\Bigr],~\label{Eq:BScurrent}
\eea
where $s_A\equiv k_A^2$, $k_A^{\mu}$ denotes the total momentum of elements in $A$. We summed over divisions $A\to A_L,A_R$, $\widetilde{A}\to\widetilde{A}_L,\widetilde{A}_R$ in (\ref{Eq:BScurrent}) so that in the first term $|A_L|=|\W A_L|$, $|A_R|=|\W A_R|$,  or in the second term  $|A_R|=|\W A_L|$, $|A_L|=|\W A_R|$. The starting point of this definition is $\phi_{a|a}=1$, $\phi_{a|b}=0$ $(a\neq b)$.  An important property is $\phi_{A|\widetilde{A}}=0$ when $A\setminus\W A\neq 0$. In other words, for nonvanishing $\phi_{A|\widetilde{A}}$ the elements of the ordered set $A$ must be indentical with those of $\widetilde{A}$. Another property is  $\phi_{A|\widetilde{A}}= \phi_{\W A|{A}}$.

\section{Forward limit of tree-level BS amplitude and one-loop BS integrand}
The colour-stripped bs amplitude with forward limit is expressed by 
summing all possible Feynman diagrams where tree-level BS subcurrents are planted to the linear propagator line between $+$ and $-$: 
\bea
&&{1\over {\ell'}^2}A^{\text{BS}}(+,\pmb\sigma,-|\,+,\pmb\rho,-)={1\over {\ell'}^2}\Sl_{\substack{\pmb\sigma\to \pmb\sigma_1|\pmb\sigma_2|...|\pmb\sigma_I\\ \pmb\rho\to \pmb\rho_1|\pmb\rho_2|...|\pmb\rho_I \\ |\pmb\sigma_i|=|\pmb\rho_i|}}\begin{minipage}{3.2cm}\includegraphics[width=3.2cm]{GenProofFeyn}\end{minipage}\\
&=&\Sl_{\substack{\pmb\sigma\to \pmb\sigma_1|\pmb\sigma_2|...|\pmb\sigma_I\\ \pmb\rho\to \pmb\rho_1|\pmb\rho_2|...|\pmb\rho_I \\ |\pmb\sigma_i|=|\pmb\rho_i|}}{1\over {\ell'}^2}\,\phi_{\pmb\sigma_1|\pmb\rho_1}\,{1\over 2 k_{\pmb\sigma_1}\cdot \ell'+k^2_{\pmb\sigma_1}}\,\phi_{\pmb\sigma_2|\pmb\rho_2}\,\dots\,{1\over 2 (k_{\pmb\sigma_1}+...+k_{\pmb\sigma_{I-1}})\cdot \ell'+(k_{\pmb\sigma_1}+...+k_{\pmb\sigma_{I-1}})^2}\,\phi_{\pmb\sigma_I|\pmb\rho_I}.\label{Eq:BSForwardLimit}
\eea
In the above expression $\ell'=\ell-k_{\pmb\sigma_1}-...-k_{\pmb\sigma_{j-1}}-k_{\pmb\sigma_{jL}}$, where we suppose $1$ belongs to 
$\pmb\sigma_j=\{\pmb\sigma_{jL},1,\pmb\sigma_{jR}\}$.

The one loop colour-stripped integrand of bs with quadratic propagators is defined by 
\bea
&&{\cal I}^{\text{BS}}(\pmb\sigma|\pmb\rho)=\Sl_{\substack{\text{CycDiv}\,\pmb\sigma\to \pmb\sigma_1,...,\pmb\sigma_I\\ \text{CycDiv}\,\pmb\rho\to \pmb\rho_1,...,\pmb\rho_I\\ |\pmb\sigma_i|=|\pmb\rho_i|}}\,\begin{minipage}{3.2cm}\includegraphics[width=3.2cm]{GenProofFeyn4}\end{minipage}\nn
&=&\Sl_{\substack{\text{CycDiv}\,\pmb\sigma\to \pmb\sigma_1,...,\pmb\sigma_I\\ \text{CycDiv}\,\pmb\rho\to \pmb\rho_1,...,\pmb\rho_I\\ |\pmb\sigma_i|=|\pmb\rho_i|}}\,{1\over {\ell'}^2}\,\phi_{\pmb\sigma_1|\pmb\rho_1}\,{1\over (\ell'+ k_{\pmb\sigma_1})^2}\,\phi_{\pmb\sigma_2|\pmb\rho_2}\,\dots\,{1\over (\ell'+k_{\pmb\sigma_1}+...+k_{\pmb\sigma_{I-1}})^2}\,\phi_{\pmb\sigma_I|\pmb\rho_I}.\label{Eq:BSQuadraticPropagatorIntegrand}
\eea
Note that the ${\cal I}^{\text{BS}}(\pmb\sigma|\pmb\rho)$ and ${\cal I}^{\text{BS}}(\pmb\rho|\pmb\sigma)$ correspond to the same BS amplitude, due to the symmetry of BS subcurrents.

Partial fraction identity induces the following identity  between tree level Feynman diagram with forward limit and one loop Feynman diagram with quadratic propagators.
\bea
{1\over {\ell'}^2}\begin{minipage}{3.2cm}\includegraphics[width=3.2cm]{GenProofFeyn}\end{minipage}+\text{cyc}(\phi_{\pmb\sigma_1|\pmb\rho_1},\phi_{\pmb\sigma_2|\pmb\rho_2},...,\phi_{\pmb\sigma_I|\pmb\rho_I})=\begin{minipage}{3.2cm}\includegraphics[width=3.2cm]{GenProofFeyn4}\end{minipage}\label{Eq:IDPartialFraction1}.
\eea
%

%
%

%
%
%
%

%
%
%
%


\begin{thebibliography}{10}


\bibitem{Bern:2008qj}
Z.~Bern, J.~J.~M. Carrasco, and H.~Johansson, {\it {New Relations for
  Gauge-Theory Amplitudes}},  {\em Phys. Rev. D} {\bf 78} (2008) 085011,
  [\href{http://arxiv.org/abs/0805.3993}{{\tt arXiv:0805.3993}}].

\bibitem{Bern:2010ue}
Z.~Bern, J.~J.~M. Carrasco, and H.~Johansson, {\it {Perturbative Quantum
  Gravity as a Double Copy of Gauge Theory}},  {\em Phys. Rev. Lett.} {\bf 105}
  (2010) 061602, [\href{http://arxiv.org/abs/1004.0476}{{\tt
  arXiv:1004.0476}}].

\bibitem{Fu:2017uzt}
C.-H. Fu, Y.-J. Du, R.~Huang, and B.~Feng, {\it {Expansion of
  Einstein-Yang-Mills Amplitude}},  {\em JHEP} {\bf 09} (2017) 021,
  [\href{http://arxiv.org/abs/1702.08158}{{\tt arXiv:1702.08158}}].

\bibitem{Chiodaroli:2017ngp}
M.~Chiodaroli, M.~Gunaydin, H.~Johansson, and R.~Roiban, {\it {Explicit
  Formulae for Yang-Mills-Einstein Amplitudes from the Double Copy}},  {\em
  JHEP} {\bf 07} (2017) 002, [\href{http://arxiv.org/abs/1703.00421}{{\tt
  arXiv:1703.00421}}].

\bibitem{Teng:2017tbo}
F.~Teng and B.~Feng, {\it {Expanding Einstein-Yang-Mills by Yang-Mills in CHY
  frame}},  {\em JHEP} {\bf 05} (2017) 075,
  [\href{http://arxiv.org/abs/1703.01269}{{\tt arXiv:1703.01269}}].

\bibitem{Du:2017kpo}
Y.-J. Du and F.~Teng, {\it {BCJ numerators from reduced Pfaffian}},  {\em JHEP}
  {\bf 04} (2017) 033, [\href{http://arxiv.org/abs/1703.05717}{{\tt
  arXiv:1703.05717}}].

\bibitem{Du:2017gnh}
Y.-J. Du, B.~Feng, and F.~Teng, {\it {Expansion of All Multitrace Tree Level
  EYM Amplitudes}},  {\em JHEP} {\bf 12} (2017) 038,
  [\href{http://arxiv.org/abs/1708.04514}{{\tt arXiv:1708.04514}}].
  
 \bibitem{Du:2025yxz}
Y.-J. Du, C.-H. Fu, Y.~Wang, and C.~Xie, {\it {Algebraic Consistency and
  Explicit Construction of One-Loop BCJ Numerators of Yang-Mills and Related
  Theories}},  \href{http://arxiv.org/abs/2511.10963}{{\tt arXiv:2511.10963}}. 
  
  \bibitem{Porkert:2022efy}
F.~Porkert and O.~Schlotterer, {\it {One-loop amplitudes in Einstein-Yang-Mills
  from forward limits}},  {\em JHEP} {\bf 02} (2023) 122,
  [\href{http://arxiv.org/abs/2201.12072}{{\tt arXiv:2201.12072}}].
  
\bibitem{ref-Kiermaier} Kiermaier, M. \textquotedblright Gravity
as the square of gauge theory, talk at Amplitudes 2010.\textquotedblright{}
Queen Mary, University of London (2010).
  

\bibitem{Bern:2010yg}
Z.~Bern, T.~Dennen, Y.-t. Huang, and M.~Kiermaier, {\it {Gravity as the Square
  of Gauge Theory}},  {\em Phys. Rev. D} {\bf 82} (2010) 065003,
  [\href{http://arxiv.org/abs/1004.0693}{{\tt arXiv:1004.0693}}].

\bibitem{Bjerrum-Bohr:2012kaa}
N.~E.~J. Bjerrum-Bohr, P.~H. Damgaard, R.~Monteiro, and D.~O'Connell, {\it
  {Algebras for Amplitudes}},  {\em JHEP} {\bf 06} (2012) 061,
  [\href{http://arxiv.org/abs/1203.0944}{{\tt arXiv:1203.0944}}].

\bibitem{Fu:2012uy}
C.-H. Fu, Y.-J. Du, and B.~Feng, {\it {An algebraic approach to BCJ
  numerators}},  {\em JHEP} {\bf 03} (2013) 050,
  [\href{http://arxiv.org/abs/1212.6168}{{\tt arXiv:1212.6168}}].

\bibitem{Fu:2014pya}
C.-H. Fu, Y.-J. Du, and B.~Feng, {\it {Note on symmetric BCJ numerator}},  {\em
  JHEP} {\bf 08} (2014) 098, [\href{http://arxiv.org/abs/1403.6262}{{\tt
  arXiv:1403.6262}}].
  
  
\bibitem{Britto:2004ap}
R.~Britto, F.~Cachazo and B.~Feng,
``New recursion relations for tree amplitudes of gluons,''
Nucl. Phys. B \textbf{715} (2005), 499-522
doi:10.1016/j.nuclphysb.2005.02.030
[arXiv:hep-th/0412308 [hep-th]].


\bibitem{Britto:2005fq}
R.~Britto, F.~Cachazo, B.~Feng and E.~Witten,
``Direct proof of tree-level recursion relation in Yang-Mills theory,''
Phys. Rev. Lett. \textbf{94} (2005), 181602
doi:10.1103/PhysRevLett.94.181602
[arXiv:hep-th/0501052 [hep-th]].
  
\bibitem{Cachazo:2013gna}
F.~Cachazo, S.~He and E.~Y.~Yuan,
``Scattering equations and Kawai-Lewellen-Tye orthogonality,''
Phys. Rev. D \textbf{90} (2014) no.6, 065001
doi:10.1103/PhysRevD.90.065001
[arXiv:1306.6575 [hep-th]].
  
  
\bibitem{Cachazo:2013hca}
F.~Cachazo, S.~He and E.~Y.~Yuan,
``Scattering of Massless Particles in Arbitrary Dimensions,''
Phys. Rev. Lett. \textbf{113} (2014) no.17, 171601
doi:10.1103/PhysRevLett.113.171601
[arXiv:1307.2199 [hep-th]].

\bibitem{Cachazo:2013iea}
F.~Cachazo, S.~He and E.~Y.~Yuan,
``Scattering of Massless Particles: Scalars, Gluons and Gravitons,''
JHEP \textbf{07} (2014), 033
doi:10.1007/JHEP07(2014)033
[arXiv:1309.0885 [hep-th]].

\bibitem{Cachazo:2014nsa}
F.~Cachazo, S.~He and E.~Y.~Yuan,
``Einstein-Yang-Mills Scattering Amplitudes From Scattering Equations,''
JHEP \textbf{01} (2015), 121
doi:10.1007/JHEP01(2015)121
[arXiv:1409.8256 [hep-th]].


\bibitem{Cachazo:2014xea}
F.~Cachazo, S.~He and E.~Y.~Yuan,
``Scattering Equations and Matrices: From Einstein To Yang-Mills, DBI and NLSM,''
JHEP \textbf{07} (2015), 149
doi:10.1007/JHEP07(2015)149
[arXiv:1412.3479 [hep-th]].

\bibitem{Mason:2013sva}
L.~Mason and D.~Skinner, {\it {Ambitwistor strings and the scattering
  equations}},  {\em JHEP} {\bf 07} (2014) 048,
  [\href{http://arxiv.org/abs/1311.2564}{{\tt arXiv:1311.2564}}].

\bibitem{He:2015yua}
S.~He and E.~Y. Yuan, {\it {One-loop Scattering Equations and Amplitudes from
  Forward Limit}},  {\em Phys. Rev. D} {\bf 92} (2015), no.~10 105004,
  [\href{http://arxiv.org/abs/1508.06027}{{\tt arXiv:1508.06027}}].

\bibitem{Cachazo:2015aol}
F.~Cachazo, S.~He, and E.~Y. Yuan, {\it {One-Loop Corrections from Higher
  Dimensional Tree Amplitudes}},  {\em JHEP} {\bf 08} (2016) 008,
  [\href{http://arxiv.org/abs/1512.05001}{{\tt arXiv:1512.05001}}].

\bibitem{He:2016mzd}
S.~He and O.~Schlotterer, {\it {New Relations for Gauge-Theory and Gravity
  Amplitudes at Loop Level}},  {\em Phys. Rev. Lett.} {\bf 118} (2017), no.~16
  161601, [\href{http://arxiv.org/abs/1612.00417}{{\tt arXiv:1612.00417}}].

\bibitem{He:2017spx}
S.~He, O.~Schlotterer, and Y.~Zhang, {\it {New BCJ representations for one-loop
  amplitudes in gauge theories and gravity}},  {\em Nucl. Phys. B} {\bf 930}
  (2018) 328--383, [\href{http://arxiv.org/abs/1706.00640}{{\tt
  arXiv:1706.00640}}].

\bibitem{Geyer:2017ela}
Y.~Geyer and R.~Monteiro, {\it {Gluons and gravitons at one loop from
  ambitwistor strings}},  {\em JHEP} {\bf 03} (2018) 068,
  [\href{http://arxiv.org/abs/1711.09923}{{\tt arXiv:1711.09923}}].



\bibitem{Cao:2025ygu}
Q.~Cao, S.~He, Y.~Zhang, and F.~Zhu, {\it {Loop-Level Double Copy Relations
  from Forward Limits}},  \href{http://arxiv.org/abs/2509.25129}{{\tt
  arXiv:2509.25129}}.



\bibitem{Xie:2024pro}
C.~Xie and Y.-J. Du, {\it {Extracting quadratic propagators by refined graphic
  rule}},  {\em JHEP} {\bf 02} (2025) 068,
  [\href{http://arxiv.org/abs/2403.03547}{{\tt arXiv:2403.03547}}].

\bibitem{Xie:2025utp}
C.~Xie and Y.-J. Du, {\it {Local vertices, quadratic propagators and
  double-copy structure of one-loop integrands from forward limits}},
  \href{http://arxiv.org/abs/2509.25632}{{\tt arXiv:2509.25632}}.

\bibitem{Kleiss:1988ne}
R.~Kleiss and H.~Kuijf, {\it {Multi - Gluon Cross-sections and Five Jet
  Production at Hadron Colliders}},  {\em Nucl. Phys. B} {\bf 312} (1989)
  616--644.

\bibitem{DelDuca:1999rs}
V.~Del~Duca, L.~J. Dixon, and F.~Maltoni, {\it {New color decompositions for
  gauge amplitudes at tree and loop level}},  {\em Nucl. Phys. B} {\bf 571}
  (2000) 51--70, [\href{http://arxiv.org/abs/hep-ph/9910563}{{\tt
  hep-ph/9910563}}].

\bibitem{Mafra:2016ltu}
C.~R. Mafra, {\it {Berends-Giele recursion for double-color-ordered
  amplitudes}},  {\em JHEP} {\bf 07} (2016) 080,
  [\href{http://arxiv.org/abs/1603.09731}{{\tt arXiv:1603.09731}}].


\end{thebibliography}

\end{document}